\documentclass[linenumbers]{aastex631}

\begin{document}

\nolinenumbers

\title{New perspective on the multiple population phenomenon in Galactic globular clusters from a wide-field photometric survey}

\author[0000-0002-1562-7557]{S. Jang}
\affiliation{Center for Galaxy Evolution Research and Department of Astronomy, Yonsei University, Seoul 03722, Korea}

\author[0000-0001-7506-930X]{A.\,P.\,Milone}
\affiliation{Dipartimento di Fisica e Astronomia ``Galileo Galilei'', Universit\`{a} di Padova, Vicolo dell'Osservatorio 3, I-35122, Padua, Italy}
\affiliation{Istituto Nazionale di Astrofisica - Osservatorio Astronomico di Padova, Vicolo dell'Osservatorio 5, IT-35122, Padua, Italy}

\author[0000-0002-1276-5487]{A.\,F.\.Marino}
\affiliation{Istituto Nazionale di Astrofisica - Osservatorio Astronomico di Padova, Vicolo dell'Osservatorio 5, IT-35122, Padua, Italy}
\affiliation{Istituto Nazionale di Astrofisica – Osservatorio Astrofisico di Arcetri, Largo Enrico Fermi, 5, Firenze 50125, Italy}

\author[0000-0002-1128-098X]{M.\,Tailo}
\affiliation{Istituto Nazionale di Astrofisica - Osservatorio Astronomico di Padova, Vicolo dell'Osservatorio 5, IT-35122, Padua, Italy}

\author[0000-0001-8415-8531]{E.\,Dondoglio} 
\affiliation{Dipartimento di Fisica e Astronomia ``Galileo Galilei'', Universit\`{a} di Padova, Vicolo dell'Osservatorio 3, I-35122, Padua, Italy}

\author[0000-0003-3153-1499]{M.\,V.\,Legnardi}
\affiliation{Dipartimento di Fisica e Astronomia ``Galileo Galilei'', Universit\`{a} di Padova, Vicolo dell'Osservatorio 3, I-35122, Padua, Italy}

\author[0000-0002-7690-7683]{G.\,Cordoni}
\affiliation{Research School of Astronomy and Astrophysics, Australian National University, Canberra, ACT 2611, Australia}

\author{T. Ziliotto}
\affiliation{Dipartimento di Fisica e Astronomia ``Galileo Galilei'', Universit\`{a} di Padova, Vicolo dell'Osservatorio 3, I-35122, Padua, Italy}

\author[0000-0003-1713-0082]{E.\,P.\,Lagioia}
\affiliation{South-Western Institute for Astronomy Research, Yunnan University, Kunming 650500, PR China}

\author[0000-0003-1757-6666]{M.\,Carlos}
\affiliation{Department of Physics and Astronomy, Uppsala University, Box 516, SE-751 20 Uppsala, Sweden}

\author{A.\,Mohandasan}
\affiliation{Dipartimento di Fisica e Astronomia ``Galileo Galilei'', Universit\`{a} di Padova, Vicolo dell'Osservatorio 3, I-35122, Padua, Italy}

\author{E. Bortolan}
\affiliation{Dipartimento di Fisica e Astronomia ``Galileo Galilei'', Universit\`{a} di Padova, Vicolo dell'Osservatorio 3, I-35122, Padua, Italy}


\author[0000-0002-2210-1238]{Y.-W.\,Lee}
\affiliation{Center for Galaxy Evolution Research and Department of Astronomy, Yonsei University, Seoul 03722, Korea}







\begin{abstract}
\nolinenumbers
Wide-field photometry of Galactic globular clusters (GCs) has been investigated to overcome limitations from the small field of view of the {\it Hubble Space Telescope} in the study of multiple populations. In particular, `chromosome maps' (ChMs) built with ground-based photometry were constructed to identify the first and second generation stars (1G and 2G) over the wide-field of view. The ChMs allow us to derive the fraction of distinct populations in an analyzed field of view. We present here the radial distribution of the 2G fraction in 29 GCs. The distributions show that all the GCs either have a flat distribution or more centrally concentrated 2G stars. Notably, we find that the fraction of 1G stars outside the half-light radius is clearly bifurcated across all mass range. It implies that a group of GCs with lower 1G fractions (hereafter Group II) have efficiently lost their 1G stars in the outermost cluster regions. In fact, in connection with the trends of the radial distribution, most GCs of Group II have spatially mixed populations, while only less massive GCs in Group I (a group with higher 1G fraction) show that feature. Lastly, we investigate links between these two groups and host cluster parameters. We find that most GCs of Group II are distributed along a broader range of galactocentric distances with smaller perigalactic distances $<$ 3.5 kpc. Besides, by using the {\it Gaia} data, it is observed that Group II GCs have higher energy on the integrals of motion diagrams than Group I GCs.

\end{abstract}





\section{Introduction} \label{sec:intro}
Observational evidence demonstrated that nearly all Galactic globular clusters (GCs) contain two main groups of stellar populations: a first population (hereafter 1G) with chemical composition similar to halo field stars, and a second population (2G) enriched in helium, nitrogen and sodium and depleted in carbon and oxygen \citep[see][and the references therein]{lee1999,gratton2004,lee2005,marino2008a,carretta2009a,renzini2015a,milone2017a,milone2020a}. In the last two decades, the multiple populations in GCs have been widely investigated, but the formation history to explain the observed abundance anomalies is a matter of ongoing debate \citep[see][for recent reviews]{bastian2018a, gratton2019a,milone2022a}. 

Some scenarios on the formation of multiple stellar populations assume that multiple populations are attributed to distinct bursts of star formation at different epochs, with 2G stars born from materials processed and ejected by 1G stars \citep[e.g.][]{decressin2007a,demink2009a,denissenkov2014a,jang2014a,jang2015,lee2015,Lee2016,dantona2016a,kim2018,renzini2022a}. One problem of these scenarios is that the enriched stars are predicted to only constitute a very small fraction of the total initial mass of clusters. Observations, however, have shown that the number fractions of 1G and 2G stars are quite similar in most GCs \citep[e.g][]{carretta2009a, milone2017a}, which is referred to as the `mass budget problem'. To address the problem, it is necessary for 1G stars to have been considerably more massive at their formation and subsequently lost a large fraction of 1G stars ($>$ 90 per cent), thereby making a significant contribution to the assembly of the Galactic halo \citep{dercole2008a,conroy2012,renzini2015a}. Alternative scenarios propose that all GC stars are coeval and that chemical enrichment of 2G stars is attributed to accretion of materials processed by massive or supermassive stars onto pre-main sequence stars \citep[e.g.][]{bastian2013a, gieles2018a}. It is also suggested that stellar mergers might be responsible for multiple populations in GCs \citep[][]{wang2020}.

Most of the multi-generation scenarios predict that 2G stars would form a more centrally segregated stellar subsystem than spatially extended 1G stars \citep[e.g.][and references therein]{dercole2008a, denissenkov2014a, dantona2016a, gieles2018a, renzini2015a}. Overtime, long-term dynamic evolution gradually diminishes the initial structural differences between 1G and 2G stars, although the strength and the progress of the evolution vary from one cluster to another. Indeed, some GCs are expected to retain memory of their initial distribution, while some can host spatially full mixed 1G and 2G stars \citep[e.g.][]{vesperini2013a, dalessandro2019a, leitinger2023}. \citet{dalessandro2019a} used {\it  Hubble Space Telescope} ({\it HST}) photometry to investigate the radial distribution of 1G and 2G stars as a function of the age/relaxation time referred to as `dynamical age'. They discovered that clusters with young dynamical ages preferentially have more centrally concentrated 2G stars, lending more support to the multi-generation scenarios.

Numerous photometric studies on multiple populations have demonstrated that this phenomenon is most efficiently identified with appropriate combinations of ultraviolet, optical, and near-infrared bands of the {\it HST} and the James Webb Space Telescope (JWST) \citep[e.g.][]{milone2012b, milone2020a,milone2023a,milone2023b, piotto2015a, niederhofer2017a, lagioia2019a, lagioia2025, dondoglio2021a,dondoglio2023, jang2021a, cordoni2022,cordoni2023,legnardi2022,legnardi2023,marino2023,mohandasan2024}.  In particular, the pseudo two-color diagram dubbed `chromosome map' (ChM) of red giant branch stars (RGBs) based on the {\it HST} photometry is well known as one of the most effective photometric tools to disentangle multiple stellar populations in GCs. The ChMs of 57 GCs constructed by \citet{milone2017a}, which were built with a combination of stellar magnitudes in the F275W, F336W, F438W, and F814W filters of the {\it HST}, allowed for an homogeneous and extensive analysis of the multiple population phenomenon with unprecedented detail.

One of the main limitations of the {\it HST} camera is its small field of view, which restricts studies on GCs and their stellar populations to the innermost cluster region. To extend the investigation to the entire cluster, it is mandatory to use wide-field photometry by means of photometric diagrams sensitive to the chemical composition of distinct stellar populations, such as photometry from suitable narrow-band filters \citep{lee2017a}, CMDs and pseudo-CMDs built with Str{\"o}emgren photometry \citep[e.g.][]{grundahl1998a, yong2008a} or appropriate combinations of the $U$, $B$, $V$ and $I$ Johnson-Cousin bands \citep[e.g.][]{marino2008a, milone2010a, milone2012a,monelli2013, dondoglio2021a,leitinger2023}. \citet{jang2022} exploited photometric diagrams of $C_{\rm U,B,I} = (U-B) - (B-I)$ pseudo color and $B-I$ color, which are sensitive to stellar populations with different light element and helium abundances along RGBs \citep{monelli2013,cordoni2020b}, to introduce ChMs from the wide-field ground-based photometry. They showed potentials of these ChMs by deriving the radial distribution of stellar populations in NGC\,288 and the fraction of 1G stars in the field of view of the ground-based photometry for NGC\,1904, NGC\,4147, NGC\,6712, NGC\,7006, and NGC\,7492.

In this paper, we extend the analysis of \citet{jang2022} to the ChMs of 29 GCs to extensively investigate the multiple population phenomenon over a wide-field of view. 
The paper is organized as follows. The data and data analysis are described in Section\,\ref{sec:data}, while Section\,\ref{sec:Radial} presents the radial distribution of the fraction of 2G stars in 29 GCs. In Section\,\ref{sec:fraction}, we derive the fraction of 1G stars in the analyzed field of view and in the region inside and outside the half-light radius from the ChMs of 29 GCs. 
Furthermore, we examine possible links between the properties we found and the GC global parameters to investigate the dependence of the multiple population phenomenon on the galactic environment. To do this, we use kinematic data and the probable merging events for Galactic GCs provided by \citet{gaiaDR2} and \citet{massari2019}. Summary and conclusions are provided in Section\,\ref{sec:summary}.





\section{Data analysis} \label{sec:data}

To investigate the multiple population phenomenon over a wide field of view, we have exploited the $\Delta c_{\rm U,B,I}$ versus $\Delta_{B,I}$ pseudo-two-color diagrams of RGB stars for 29 GCs from \citet[][]{jang2022}. In their work, they analyzed the state-of-art photometry and astrometry of 43 GCs provided by \citet{stetson2019} for 43 Galactic GCs. For each cluster, they identified probable cluster members by combining photometry with stellar proper motions and parallax measurements from Gaia eDR3 \citep{gaia2021a}. Additionally, they derived high-resolution reddening maps and corrected the photometry for differential reddening if necessary. The $\Delta c_{\rm U,B,I}$ vs $\Delta_{B,I}$ ChMs were derived from the {\it I} versus C$_{\rm U,B,I}$ pseudo-CMD and the {\it I} versus {\it B-I} CMD, which maximize the separation between distinct stellar populations. 
For more information about the ChMs of 29 Galactic GCs, the reader may refer to \citet{jang2022}.

\begin{figure*}
\begin{center}
\includegraphics[height=8cm,trim={0cm 0cm 0cm 11cm},clip]{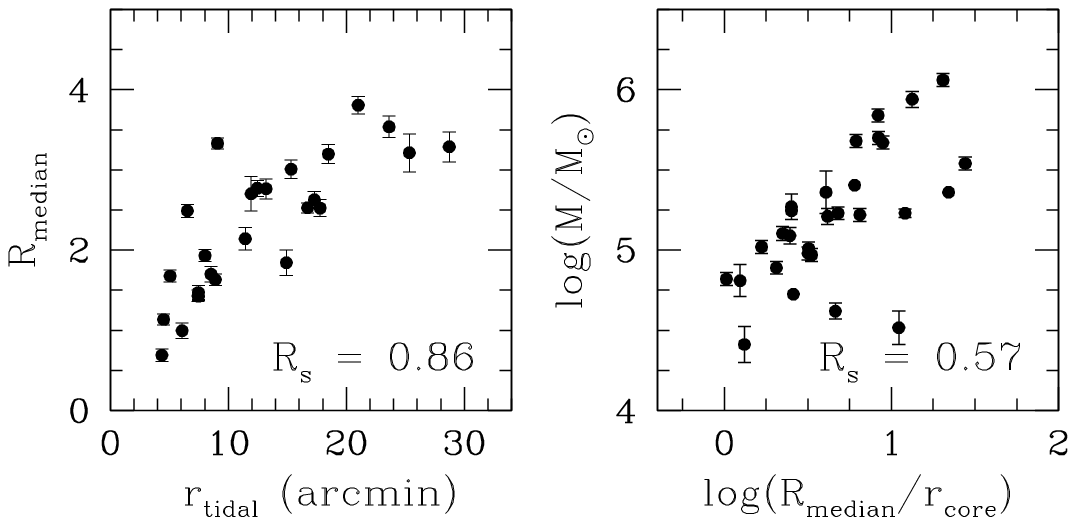}

\caption{R$_{\rm median}$ against the tidal radius of the cluster \citep[from the 2010 version of the][catalog]{harris1996a}. The Spearman’s rank correlation coefficients (R$_{\rm s}$) are reported in each panel. }
    \label{fig:Rmed}
\end{center}
\end{figure*}

Although the ground-based catalogues cover almost the whole area of each cluster, higher stellar density in the central region causes blending to affect the photometry of stars. Thus, a number of stars located in the central region are removed in the step taken to select a sample of well-measured stars before separating stellar populations. Indeed, 96\% and 81\% of the analyzed stars in 29 GCs are located outside their core and half-light radii, respectively, which implies that the results from this study are more representative of the region outside of the central region of clusters. In this context, we found a strong correlation between the median radius of the analyzed stars, R$_{\rm median}$, and the tidal radius of clusters taken from \citet[2010 version of the][catalog]{harris1996a}, r$_{\rm tidal}$, as shown in the left-hand panel of Figure\,\ref{fig:Rmed}. This is confirmed by the Spearman’s rank correlation coefficient, R$_{\rm s}$ = 0.86, which is significant at the {\it p}-value $<$ 0.00001. Thus, the cluster mass correlates with $Log(R_{\rm median}/r_{\rm core})$ as shown in the right-hand panel, which is expected from the well-known relation between the cluster mass and the King-model central concentration defined as $Log(r_{\rm tidal}/r_{\rm core})$. The uncertainty associated with R$_{\rm median}$ is calculated by bootstrapping with replacement over the sample of RGB stars, then repeated 1000 times. We adopted this method due to its suitability for data with small sample sizes or unknown distributions. We considered one standard deviation of the bootstrapped measurements as an error estimate. 







\begin{figure*}
\begin{center}
	\includegraphics[height=12cm,trim={0cm 7.3cm 0cm 0cm}]{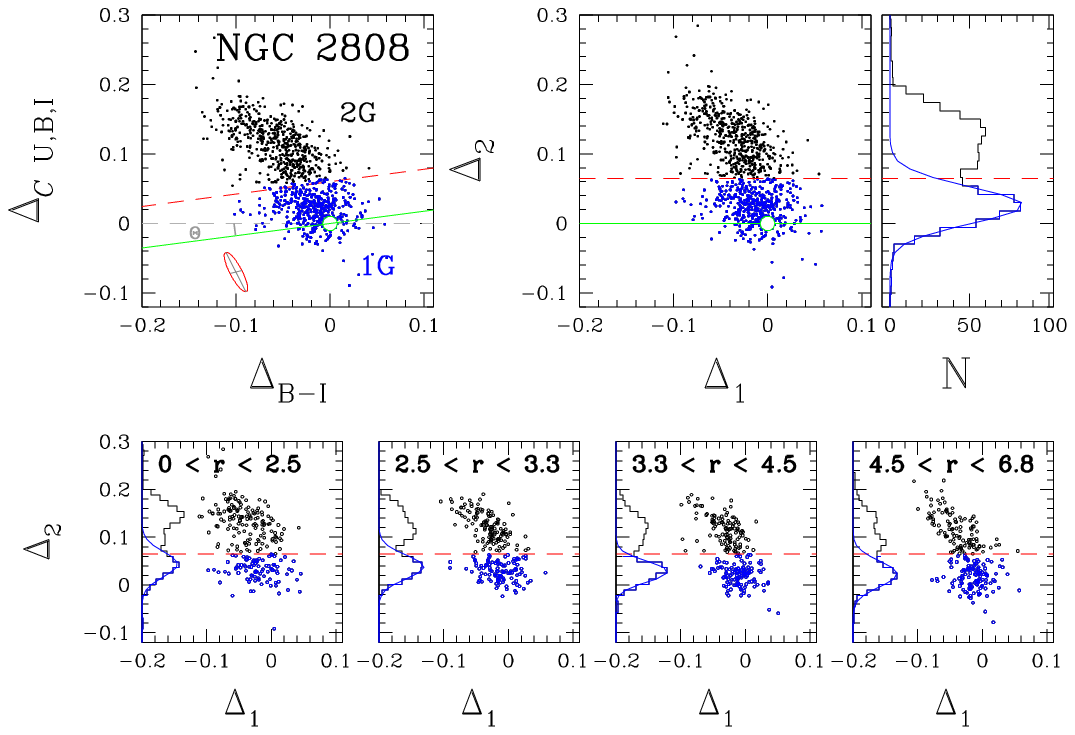}
    \caption{\textit{Upper panels:} the figure illustrates the procedure used to identify 1G and 2G stars in NGC\,2808 and derive the number fraction of each population. The top left panel shows the wide-field ground based ChM $\Delta c_{\rm U,B,I}$ versus $\Delta_{\rm B,I}$ in NGC\,2808, which is adopted from \citet{jang2022}. The red-dashed line drawn by eye separates the selected 1G and 2G stars, which are colored blue and black, respectively. The green line has the same angle $\Theta$ as the red-dashed line with respect to the horizontal line but goes through the frame's origin. The new coordinates $\Delta_{2}$ versus $\Delta_{1}$ in the top middle panel are reproduced by rotating clockwise by the angle $\Theta$ the plot in the top left-hand panel. The histogram in the top right panel shows the distributions of the $\Delta_{2}$ values. The red ellipse in the top left panel shows the distribution of the observational error. \textit{Bottom panels:} Same as the upper middle and right panels but at the different radial intervals from the cluster center. }
    
    \label{fig:nf}
\end{center}
\end{figure*}

Figure\,\ref{fig:nf} illustrates the procedure to derive the fraction of 1G stars in NGC\,2808 from its ChM. We mostly followed the method by \citet[][see their Figure 8]{milone2017a}, which was applied on the $\Delta c_{\rm F275W,F336W,F814W}$ versus $\Delta_{\rm F275W,F814W}$ diagram. \citet{jang2022} noticed that the slope of the 1G stars in the ground-based ChMs of GCs with extended 1G sequences seems to vary from cluster to cluster, which have not been observed in the ChMs based on {\it HST} photometry. 
For obtaining a new coordinate system where 1G and 2G stars could be easily identified, we defined by eye a gap between 1G and 2G stars with a line (see the dashed red line in the top left panel). We then took as 1G stars below the red line, while we defined the remaining stars as 2G stars in all the panels, which are colored blue and black, respectively. 
The red ellipse shown in the top left panel indicates the expected distributions of the observational errors and includes 68.27 per cent of the simulated stars. 
The green line has the same angle $\Theta$ as the dashed red line, with respect to the horizontal line, but it goes through the origin of the frame. The new coordinate $\Delta_{2}$ versus $\Delta_{1}$ in the top middle panel is generated by rotating the top left panel diagram by the angle $\Theta$ around the origin of the reference frame. The histogram plotted in the top right panel represents the $\Delta_{2}$ distribution of stars in the cluster. 
To estimate the fraction of 1G stars with respect to the total number of stars, the Gaussian function is fitted to the histogram distribution of the selected 1G stars (blue continuous line in the top right panel of Figure\,\ref{fig:nf}), and we derive the fraction as the ratio between the area under the Gaussian and the total number of stars in the ChM. 
Bottom panels of Figure\,\ref{fig:nf} illustrate the same procedure described above, but performed with the groups of stars at different radial intervals from the cluster center to derive the radial distribution of the 1G fraction. The radial intervals indicated in the bottom panels are determined to include the same number of stars in each interval. The resulting radial distribution of the fraction of 2G stars in NGC\,2808 is illustrated in one panel of Figure\,\ref{fig:RD1nero}. 

As described above, the analyses presented in this paper are mostly limited to stars in the outermost cluster regions. Numerous studies have found that there are differences in internal dynamics and radial distributions between 1G and 2G stars in many clusters \citep[e.g.][]{Libralato2023,cordoni2024}. Therefore, the criteria for our selection pose limitations in interpreting the results and understanding the properties of the entire cluster. In that regard, using the {\it HST} data for the inner cluster region in combination with the ground-based data provides a significant advantage in performing a homogeneous analysis of the wide-field spatial extent of a large sample of GCs \citep[e.g.][]{leitinger2023}. \citet{milone2017a} documented the fractions of 1G stars in Galactic GCs, which were derived from the {\it HST}-based ChMs of RGB stars. In this paper, we used these 1G fractions by \citet{milone2017a}  for the inner cluster region, together with our results, to investigate multiple populations over a wide field of view.
We note that stars excluded during our photometric cleaning, along with cluster regions without observed stars caused by the limitations of the field of view for some clusters, might introduce uncertainties in the number density profile as a function of radius, potentially leading to subtle biases in our results.

\section{Radial distribution of the fraction of 2G stars in 29 GCs} \label{sec:Radial}



\begin{figure*}
\begin{center}
	\includegraphics[height=18.5cm,trim={0cm 0cm 0cm 0cm}]{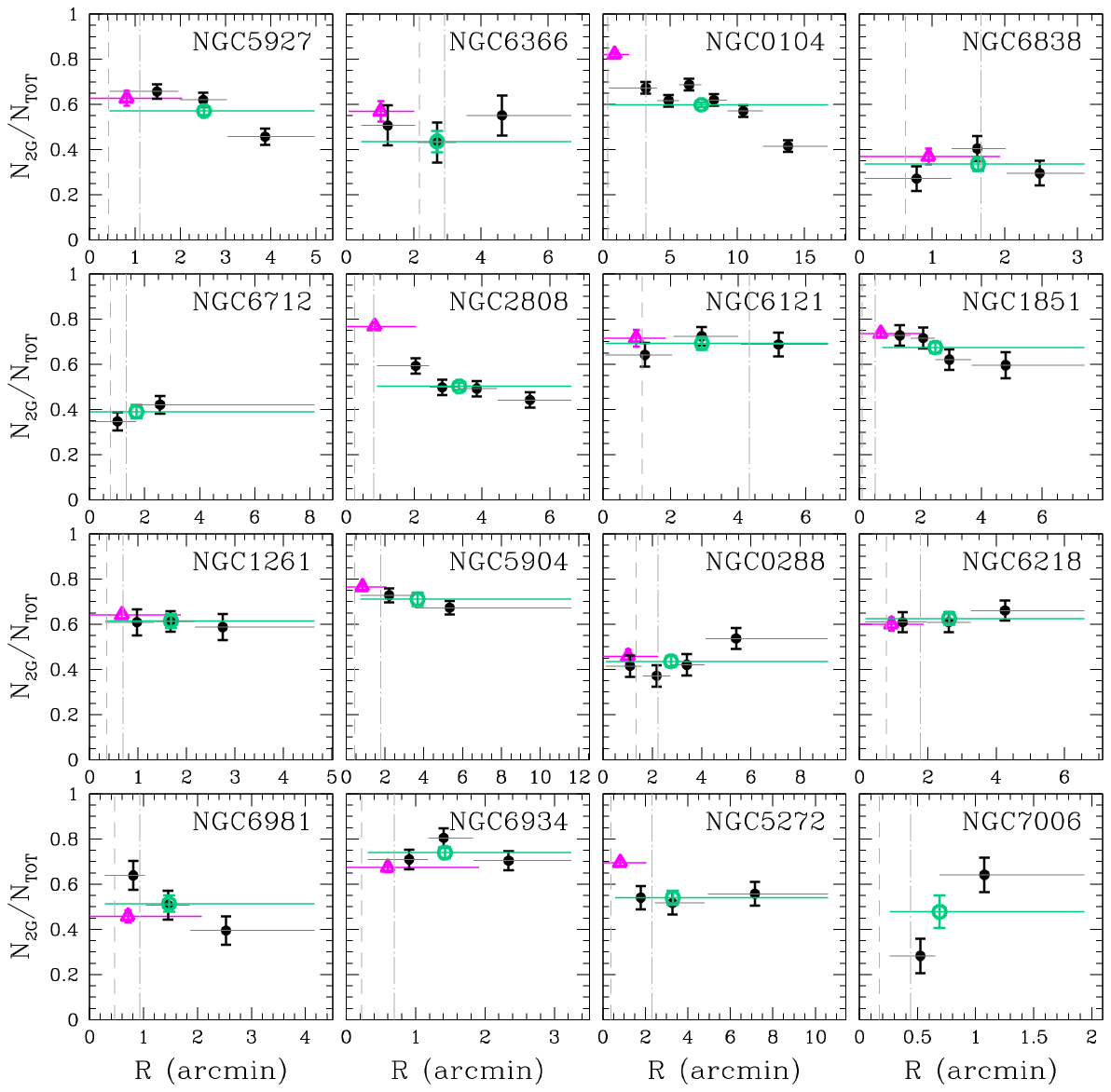}
    \caption{Fraction of 2G stars as a function of radial distance for NGC\,5927, NGC\,6366, NGC\,104, NGC\,6838, NGC\,6712, NGC\,2808, NGC\,6121, NGC\,1851, NGC\,1261, NGC\,5904, NGC\,288, NGC\,6218, NGC\,6981, NGC\,6934, NGC\,5272, and NGC\,7006. The clusters are sorted according to their metallicity, from the most metal-rich to the most metal-poor. Black circles mark the results derived from ground-based ChMs of GCs, whereas the aqua circle indicates the 2G fraction in the analyzed entire field of view. The magenta triangle indicates the fraction of 2G stars from \citet[][]{milone2017a,milone2020a} based on the {\it HST} photometry. Horizontal lines mark the extension of each radial interval. The vertical dash and dashed-dotted lines indicate the core and the half-light radius.}
    \label{fig:RD1nero}
\end{center}
\end{figure*}

\begin{figure*}
\begin{center}
	\includegraphics[height=18.5cm,trim={0cm 0cm 0cm 0cm}]{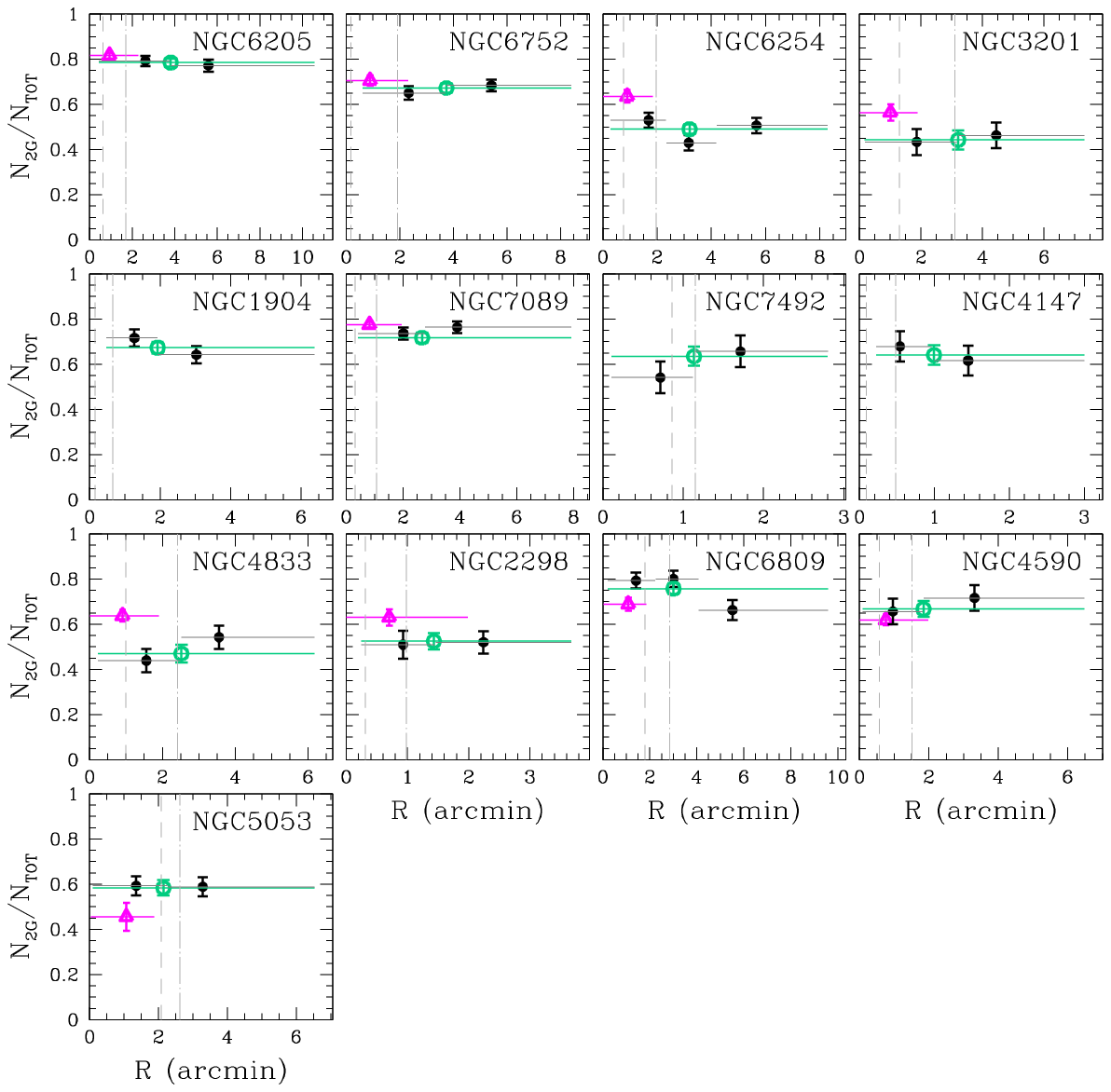}
    \caption{As in Figure\,\ref{fig:RD1nero}, but for NGC\,6205, NGC\,6752, NGC\,6252, NGC\,3201, NGC\,1904, NGC\,7089, NGC\,7492, NGC\,4147, NGC\,4833, NGC\,2298, NGC\,6809, NGC\,4590, and NGC\,5053.}
    \label{fig:RD2nero}
\end{center}
\end{figure*}

Figure\,\ref{fig:RD1nero} and Figure\,\ref{fig:RD2nero} illustrate a collection of the radial distribution of the number fraction of 2G stars in 29 GCs as a function of the median radius of the stars within each radial interval. We sorted these clusters according by metallicity, from the most metal-rich (NGC\,5927, [Fe/H]=$-$0.49) to the most metal-poor (NGC\,5053, [Fe/H]=$-$2.27). The relevant results are indicated with black dots, and the gray horizontal lines mark the extension of each radial interval determined to include the same number of stars. The 2G fraction of the entire analyzed stars is marked with the aqua circle at their median radius. In Table\,\ref {tab:DR}, we provide this median radius of our entire sample for each cluster in units of half-light radius, R$_{\rm median}$/r$_{\rm hl}$, together with the corresponding fraction, to show in which region of the cluster the median radius of our entire samples is located.
The magenta triangle indicates the literature results from \citet{milone2017a} based on the {\it HST} photometry, which is not indicated when the photometry is not available. 
The error associated to the 2G fraction has been determined by bootstrapping statistics. We generated random numbers as an equal size sample of the analyzed stars, then measured a ratio of the numbers smaller than the 2G fraction, then this procedure was repeated 1000 times. The derived errors refer to one standard deviation of the bootstrapped measurement.   

Corresponding to what is expected in most formation scenarios, in the majority of GCs, the 2G fraction in the {\it HST} field of view is generally higher than that in the analyzed field of view of the ground-based photometry (see magenta and aqua dots in Figure\,\ref{fig:RD1nero} and Figure\,\ref{fig:RD2nero}). At the same time, we detected a large degree of variety in the 2G fraction as a function of the radial distance from the cluster center.
Based on our results shown in Figure\,\ref{fig:RD1nero} and Figure\,\ref{fig:RD2nero}, we classified them into two groups according to the radial variation with its uncertainty. For metal-rich clusters ([Fe/H] $<$ -1.6, more metal-rich than NGC\,7089) for which color separation between 1G and 2G in the ChM tends to be relatively clear, the classification exploited also the literature results by \citet{milone2017a}.
We identified the centrally concentrated 2G stars in NGC\,5927, NGC\,104, NGC\,2808, NGC\,1851, NGC\,6981, NGC\,5272, NGC\,6254, NGC\,3201, and NGC\,6809, whereas stellar populations are spatially mixed in NGC\,6366, NGC\,6838, NGC\,6712, NGC\,6121, NGC\,1261, NGC\,5904, NGC\,288, NGC\,6218, NGC\,6934, NGC\,6205, NGC\,6752, NGC\,1904, NGC\,7089, NGC\,7492, NGC\,4147, NGC\,4833, NGC\,2298, NGC\,4590, and NGC\,5053. 
NGC\,7006, where we observe a larger fraction of 2G stars in the external region, is a possible exception. However, the smallest sample size restricts us from making a firm conclusion for this cluster.

\citet{dondoglio2021a} investigated stellar populations along the horizontal branch (HB) by means of the ground-based photometric diagram, V versus C$_{\rm U,B,I}$ of 4 metal-rich Galactic GCs, NGC\,104, NGC\,5927, NGC\,6366, and NGC\,6838. They derived the radial distribution of 2G fraction from distinct sequences of 1G and 2G stars on the red HB, which can be detected in metal-rich GCs with [Fe/H] $\gtrsim$ -1.0 with appropriate photometric diagrams. Our finding is consistent with their conclusion that 2G stars in NGC\,104 and NGC\,5927 are more centrally concentrated than 1G stars, while the distribution is quite flat in NGC\,6366, and NGC\,6838. 

\citet{leitinger2023} performed wide view analysis of multiple populations in 28 Galactic GCs by combining {\it HST} photometry with the wide-field ground-based photometry. 
For the 10 dynamically young GCs (age/relaxation time $<$ 4.5), for which they provided radial distributions of the fraction of 2G stars, we were able to compare their results with ours. We noticed that similar radial distributions to ours are found in NGC\,288, NGC\,2808, NGC\,4590, NGC\,5272, NGC\,5904, NGC\,6205, NGC\,7089, NGC\,6809, and NGC\,5053. In contrast, they found that 1G stars in NGC\,3201 is more centrally concentrated than 2G stars at odds with our result of centrally concentrated 2G stars in this cluster. 

Most previous studies qualitatively corroborated a 2G concentration in NGC\,3201 \citep{carretta2010,Kravtsov2010,lucatello2015,kamann2020a,Kravtsov2021}, but recent studies on the radial configuration of stellar populations in NGC\,3201 have presented somewhat conflicting results.
\citet{hartmann2022a} discovered a 1G central concentration by combining {\it HST} photometry with photometry from the S-PLUS survey, while \citet{mehta2025} found that more centrally concentrated 2G stars by introducing new photometric bands sensitive to the different chemical compositions of multiple populations from {\it Gaia} DR3 low-resolution XP spectra. 
Fuzzy separation of stellar populations in the photometric diagrams, along with potential incompleteness in the correction for severe differential reddening, could be one of the reasons for this discrepancy observed in NGC\,3201. Improved photometric and spectroscopic data would be required to reach a consistent result on the radial distribution of stellar populations in NGC\,3201.

\section{The fraction of 1G stars and global cluster parameters } \label{sec:fraction}
  
\begin{figure*}
\begin{center}

\includegraphics[height=12cm,trim={0cm 0cm 0cm 0cm},clip]{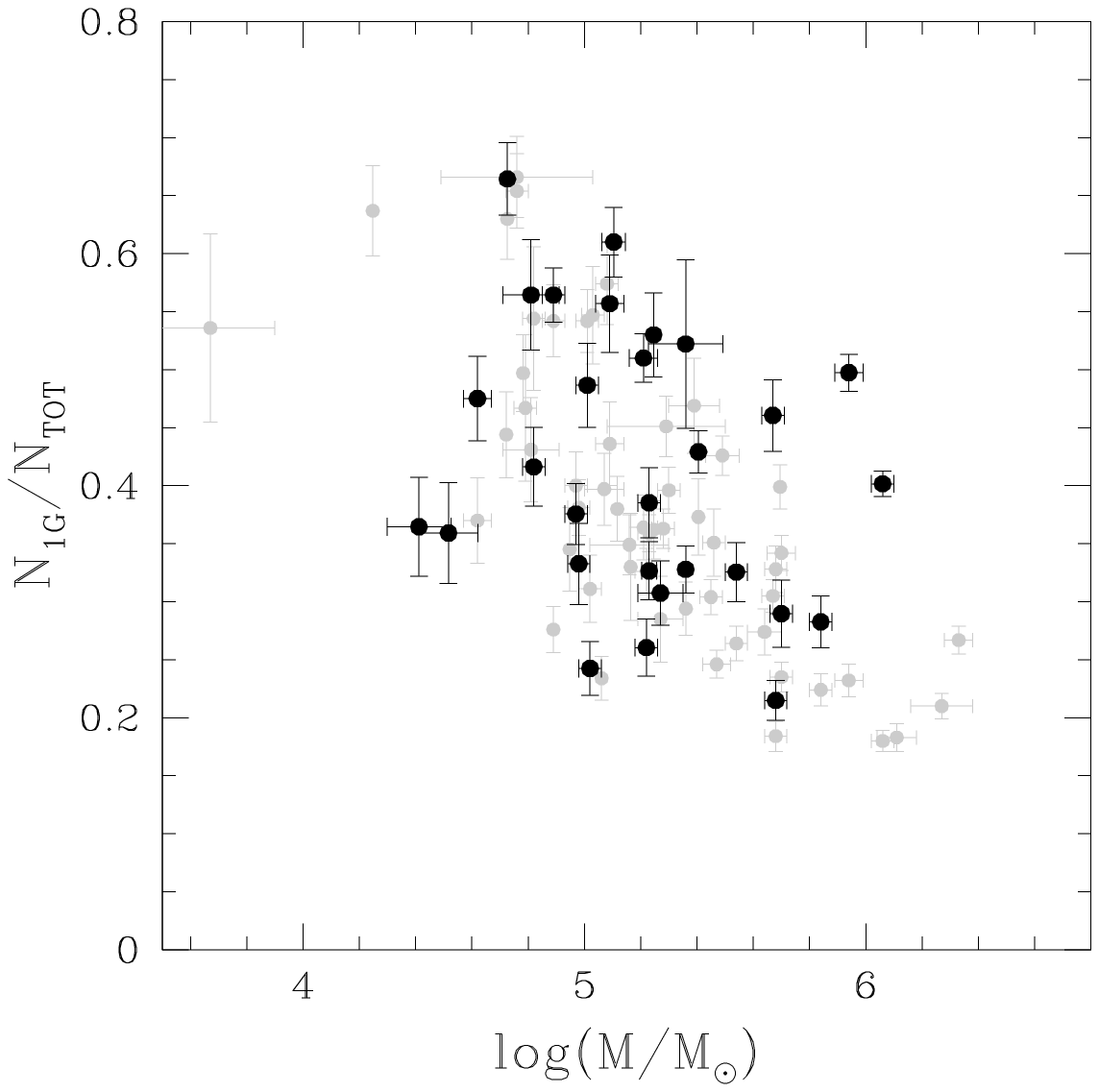}

\caption{Fraction of 1G stars against the present-day cluster mass for Galactic GCs. Gray dots indicate literature results, which were derived from the {\it HST}-based ChMs \citep[][]{milone2017a, dondoglio2021a}, whereas the fraction calculated from the ground-based ChMs are marked with black points. Cluster masses are from \citet[][]{baumgardt2018a}.}
\label{fig:f1stetson}
\end{center}
\end{figure*}

In the following, we investigate the fraction of 1G stars in the analyzed field of view. As described in Section\,\ref{sec:data}, the procedure to estimate the fraction with respect to the total number of stars (N$_{\rm TOT}$) is illustrated in the upper panels of Figure\,\ref{fig:nf} for NGC\,2808. 
The resulting fractions of 1G stars are listed in Table\,\ref {tab:DR}, where we also provide the total number of RGB stars and the median radial distance of stars included in the $\Delta_{C \rm U,B,I}$ vs.\,$\Delta_{\rm B,I}$ ChMs in units of half-light radius, as mentioned in Section\,\ref{sec:Radial}.
The fractions of 1G stars against the present-day cluster mass for Galactic GCs are plotted in Figure\,\ref{fig:f1stetson}. It is immediately clear that the 1G fraction generally anticorrelates with cluster mass as already reported in the literature, but, at the same time, it is clearly bifurcated across all mass ranges. 

\begin{figure*}
\begin{center}
\includegraphics[height=10cm,trim={0cm 0cm 0cm 5.6cm},clip]{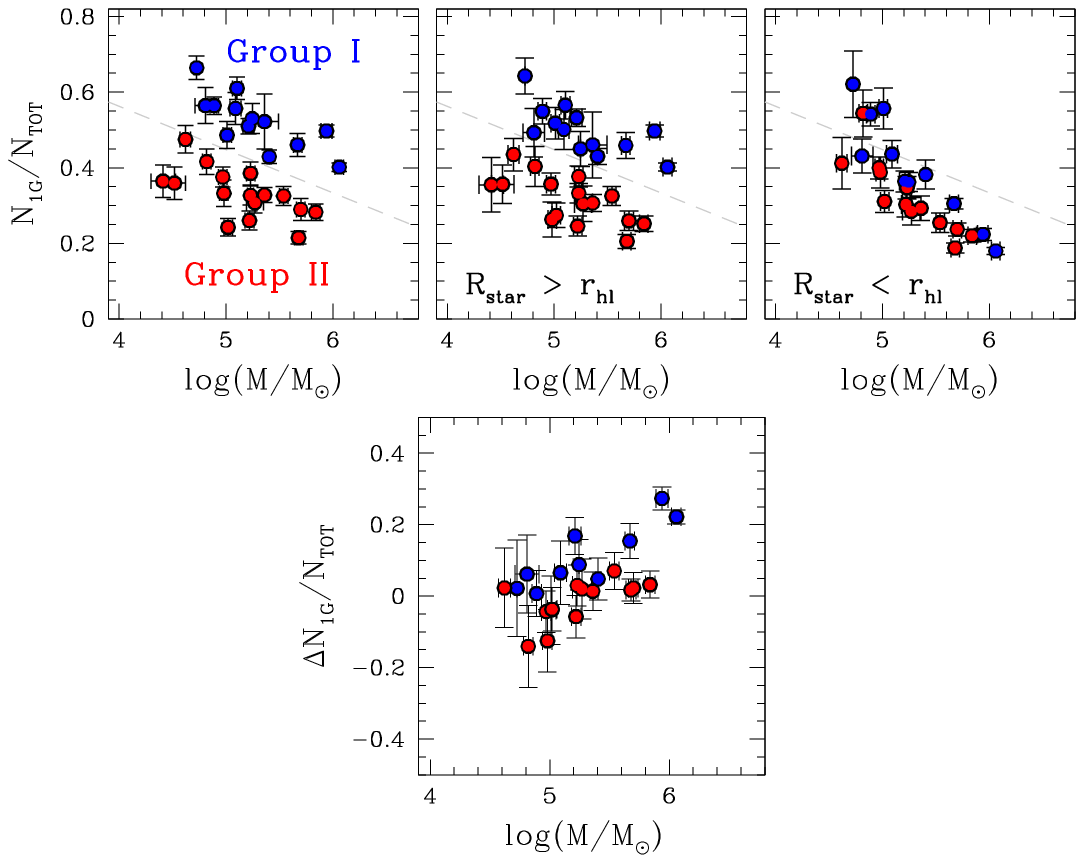}

\caption{The top left panel is the same as Figure \ref{fig:f1stetson}, but we divided the investigated targets into two groups, namely Group I (blue dots) and Group II (red dots). The top middle panel show the 1G fraction in the region outside the half-light radius as a function of cluster mass. The top right panel illustrates the fraction of 1G stars inside the half-light radius, which is derived from the {\it HST}-based ChMs. The difference in the fractions of 1G stars inside and outside the half-light radius is plotted as a function of the cluster mass in the bottom panel. }
\label{fig:twogroup}
\end{center}
\end{figure*}

As shown in the top left panel of Figure\,\ref{fig:twogroup}, we divided them into two groups of GCs based on the visually identifiable separation, Group I with higher 1G fraction and Group II with lower 1G fraction at a given mass. The Spearman's rank correlation coefficient, R$_{\rm s}$ for the relation between N$_{\rm 1G}$/N$_{\rm TOT}$ and the cluster mass are -0.32, -0.79 and -0.67 for the whole sample of GCs, Group I and II, which are significant at the p-value = 0.047, 0.001, and 0.002, respectively.
Although we showed in Figure\,\ref{fig:Rmed} that our sample of stars represents the region outside of the central region of clusters, for being more consistent in comparative studies among GCs, we also derived the fraction of 1G stars in the region outside the half-light radius, which is measured by excluding stars within the half-light radius. This fraction is also listed in Table\,\ref {tab:DR} and is shown in the top middle panel of Figure\,\ref{fig:twogroup}. As expected, despite of a larger uncertainty due to a smaller number of analyzed stars, the dichotomy still exists when removing stars within the half-light radius. For comparison, we also derived the fraction of 1G stars inside the half-light radius, as shown in the top right panel. This is measured by removing stars outside the half-light radius from the {\it HST}-based ChM \citep[][]{milone2017a}. 
A small difference in the fractions between Group I and II clusters across cluster mass ranges is also detected, but the two groups are not separated on the diagram.   
 
 Results from the fractions of 1G stars inside and outside the half-light radii imply that clusters in Group II may have more efficiently lost their 1G stars in the outermost cluster regions, while Group I GCs have relatively retained the memory of the initial distribution of 1G and 2G stars. This is further supported by the fact that in Group I, massive clusters have a significantly higher fraction of 1G stars outside the half-light radius compared to the fraction of 1G stars inside the half-light radius. The bottom panel illustrates the difference in the 1G fractions between the regions outside and inside the half-light radius, $\Delta N_{\rm 1G}/N_{\rm TOT}$, which is obtained by subtracting the 1G fraction in the region inside the half-light radius from the fraction in the region out of the half-light radius. This difference implies the extent of central concentration of stellar populations, showing the correlation with cluster mass. However, clusters of Group II, which are expected to be relaxed and have their populations relatively mixed, have lower $\Delta N_{\rm 1G}/N_{\rm TOT}$ than those of Group I. 

\subsection{Relation with the radial distribution of the fraction of 2G stars in GCs}\label{sec:relation1}

\begin{figure*}
\begin{center}
\includegraphics[height=9cm,trim={0cm 0cm 0cm 0cm},clip]{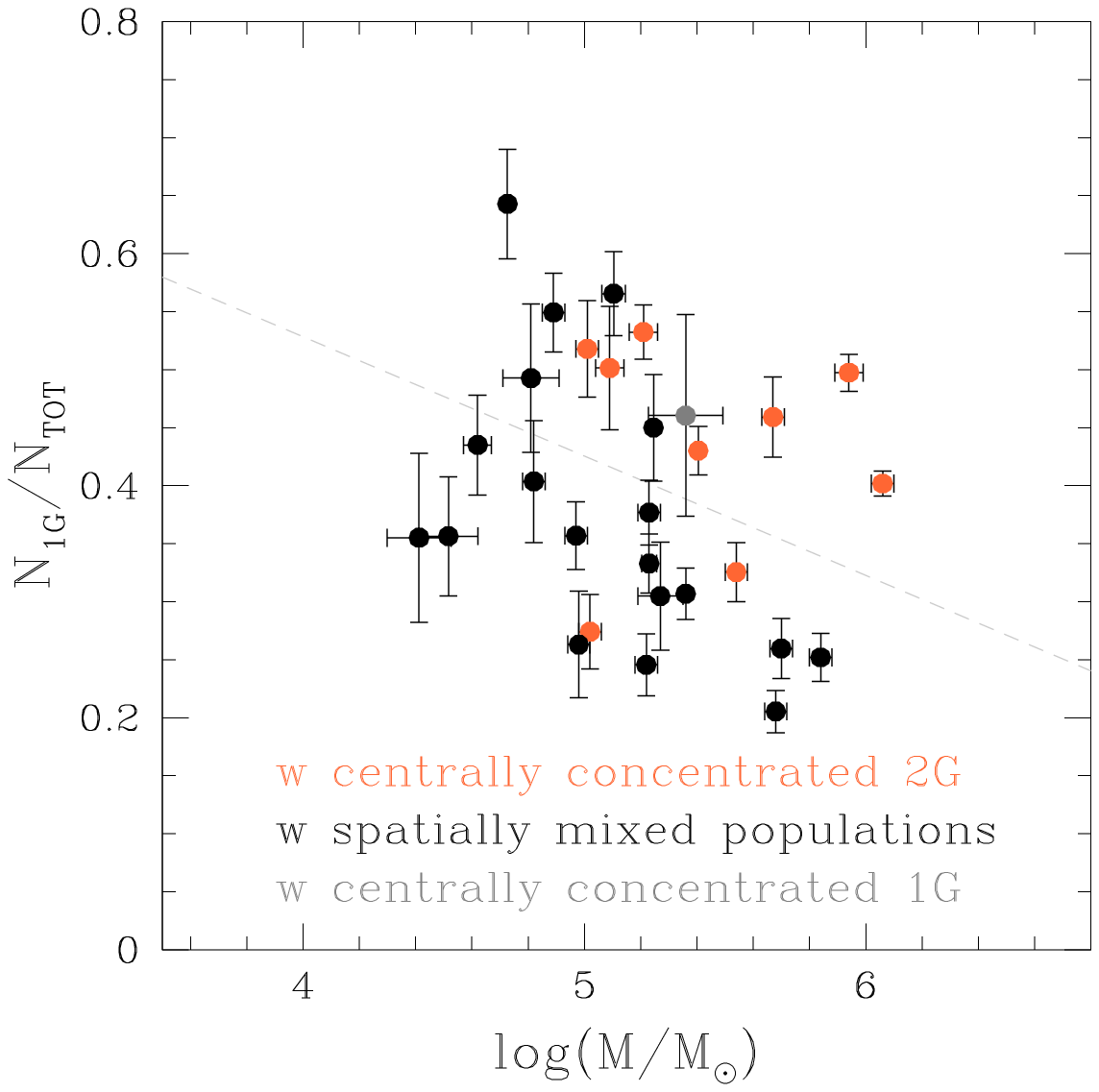}

\caption{Same as the middle panel of Figure\,\ref{fig:twogroup}, but color-coded by classification by the trend of radial distribution. Black dots mark the clusters with spatially mixed stellar populations, while clusters with centrally concentrated 2G stars are colored in orange. 
Gray points indicate NGC\,7006 with centrally concentrated 1G stars.}
\label{fig:relation1}
\end{center}
\end{figure*}

Figure\,\ref{fig:relation1} is the same as the middle panel of Figure\,\ref{fig:twogroup}, but color-coded according to the trends of radial distribution classified in Section\,\ref{sec:Radial}. We found that nearly all clusters of Group II have spatially mixed stellar populations. Two possible exception are NGC\,1851 and NGC\,6809 with centrally concentrated 2G stars. NGC\,1851 is one of the peculiar GCs having additional stellar populations with enhancement in Fe and heavy elements. This population forms the additional sequence on the ChM, which is considered as 2G stars for simplicity when deriving the fraction of stellar populations \citep{jang2022}. It therefore would become somewhat difficult to directly compare the fraction of this cluster with those of other clusters. In spite of 2G stars more centrally concentrated in NGC\,6809, it is known for this cluster to have undergone a significant amount of mass loss of M$_{\rm current}$/M$_{\rm initial}$ = 0.261 \citep{baumgardt2019,leitinger2023}, as predicted for Group II GCs. Thus, we could not clearly conclude that the centrally concentrated 2G stars in NGC\,6809 reflects its initial spatial distribution of populations.  
Conversely, the radial distributions of 2G stars in Group I GCs are more complex. Massive GCs of Group I have more centrally concentrated 2G stars, while the radial distributions in less massive Group I GCs appear to be flat. This is hardly surprising, as less massive clusters tend to have shorter relaxation time scales \citep{Spitzer1987}. It is expected that Group II GCs have undergone much dynamical mixing, all having spatially blended stellar populations, while relatively mild dynamic evolution has proceeded in Group I GCs, retaining their initial spatial condition, at least, in massive GCs. 



As described above, \citet{leitinger2023} combined {\it HST} photometry with the wide-field ground-based photometry to investigate the multiple population phenomenon over a wide-field of view. The sample of GCs investigated by \citet{leitinger2023} comprises 21 clusters, which are also studied in this paper. 
While the radial distribution of each population intuitively illustrates the changes in the number fraction of each population as a function of radius, quantifying the difference in the radial distribution between 1G and 2G stars is rather challenging. In contrast, the cumulative radial distribution of the stars in each population, which \citet{leitinger2023} primarily used, schematically displays the differences in the distribution of 1G and 2G stars within the cluster, and the difference can be quantified by the parameter $A^{+}$ introduced by \citet{Alessandrini2016}. 
They quantified different radial profiles of 1G and 2G stars by using the area enclosed between their cumulative radial distributions, $A^{+}$, across the full extent of each cluster. This parameter indicates if a cluster has centrally concentrated 2G stars ($A^{+} <$ 0), concentrated 1G stars ($A^{+} >$ 0), or spatially mixed stellar populations ($A^{+} \sim$ 0). They investigated the resulting $A^{+}$ as a function of dynamical age, which is defined as the ratio of a cluster’s age to its half-mass relaxation time, age/t$_{\rm rh}$. They found that dynamically old GCs (age/t$_{\rm rh} >$ 4.5) all have $A^{+} \sim$ 0, while a wide range of $A^{+}$ is observed in dynamically young GCs (age/t$_{\rm rh} <$ 4.5).

\begin{figure*}
\begin{center}
\includegraphics[height=8cm,trim={0cm 0cm 0cm 9cm},clip]{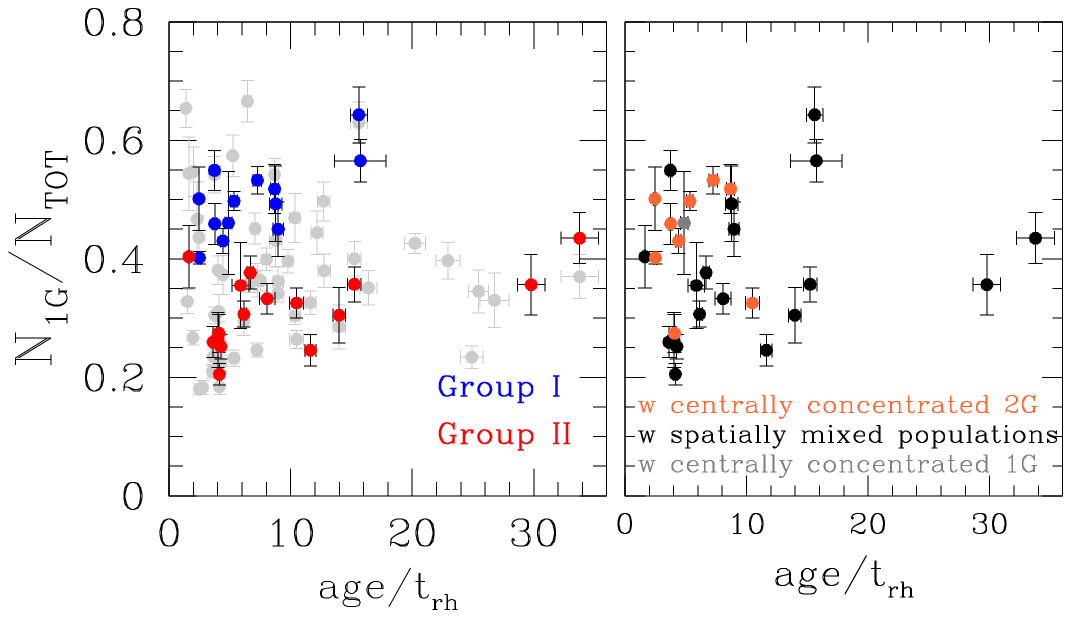}

\caption{Fraction of 1G stars as a function of dynamical age, which is defined as the ratio of the cluster’s age to its half-mass relaxation timescale. Gray dots plotted in the left panel represent literature results derived from the {\it HST}-based ChMs \citep{milone2017a,dondoglio2021a}. Ages and half-mass relaxation time, t$_{\rm rh}$, are taken from \citet{kruijssen2019} and \citet{Baumgardt2017}, respectively. Clusters are color-coded according to the group of GCs (left panel) and the categorized trend of radial distributions of 2G fraction (right panel).}
\label{fig:relation2}
\end{center}
\end{figure*}

Figure\,\ref{fig:relation2} illustrates the fraction of 1G stars as a function of dynamical age. The fractions of 1G stars in Group I and Group II GCs appear to weakly to moderately correlate with dynamical age. The Spearman’s rank correlation coefficients are R$_{\rm s}$ = 0.42 with a p-value = 0.074 for Group I and 0.38 with a p-value = 0.074 for Group II, showing slight but ultimately inconclusive correlations.
In addition, we noticed that the 1G fractions, including the literature result represented by gray points, of dynamically old GCs appear to exhibit relatively similar values. In contrast, dynamically young GCs display a wide range of 1G fractions. Considering the result of \citet{leitinger2023} that dynamically old clusters all have fully mixed populations, this implies that such clusters have undergone significant dynamical mixing during their lifetime and lost their 1G stars preferentially in their outer regions, leading to similar 1G fractions.
The right-hand panel is the same as the left-hand panel, but color-coded according to the categorized trend of radial distributions of 2G fraction. We noticed that most dynamically old clusters (age/t$_{\rm rh} \gtrsim$ 10) have spatially full mixed stellar populations, whereas clusters with lower dynamical age (age/t$_{\rm rh} \lesssim$ 10) have various spatial configurations of stellar populations, including not only centrally concentrated 2G, but also spatially blended stellar populations. This result is qualitatively in agreement with previous findings by \citet{leitinger2023}.

\subsection{Relation with global dynamical properties}\label{sec:relation1}

\begin{figure*}
\begin{center}
\includegraphics[height=8cm,trim={0cm 0cm 0cm 10cm},clip]{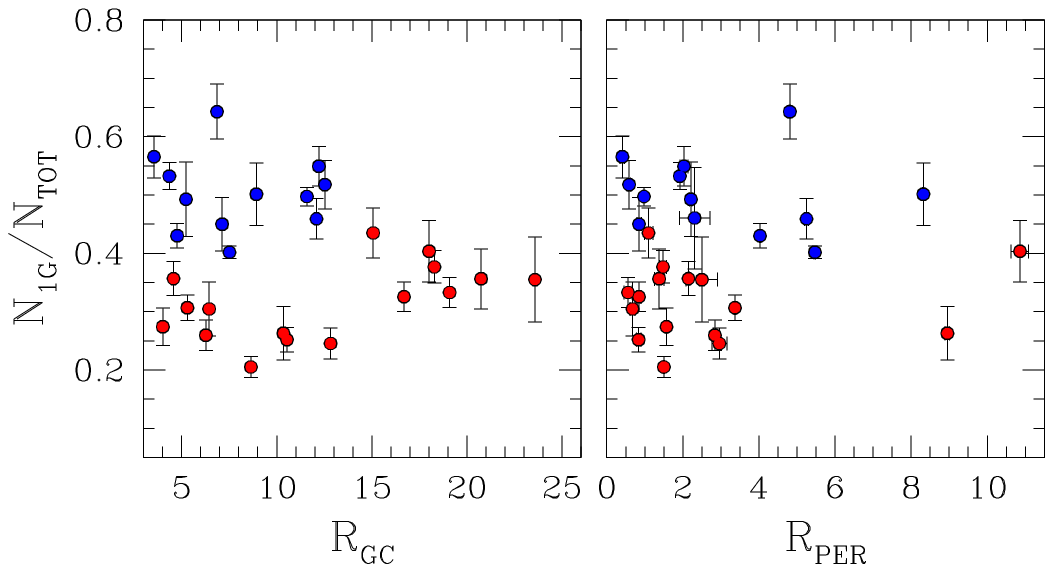}

\caption{The fraction of 1G stars as a function of galactic and perigalactic radius. The Group I and Group II GCs defined in the N$_{\rm 1G}$/N$_{\rm TOT}$ and cluster mass plane are represented with blue and red circles, respectively.}
\label{fig:peri}
\end{center}
\end{figure*}

Here, we investigate relations between the two groups of GCs and kinematic properties of the host GCs. Figure\,\ref{fig:peri} shows the 1G fraction as a function of galactic radius, $R_{\rm GC}$, and perigalactic distance, $R_{\rm PER}$ \citep[from][]{baumgardt2018a}, which reveals that the Group II GCs in red tend to have a wide range of $R_{\rm GC}$ and smaller perigalactic radii $R_{\rm PER}$ $<$ 3.5 kpc, except for most metal-poor GCs, NGC\,4590 and NGC\,5053 with the largest perigalactic radii. This is consistent with previous findings by \citet{zennaro2019a,milone2020a} that GCs with large perigalactic distances ($R_{\rm PER}$ $>$ 3.5 kpc) tend to have larger fraction of 1G stars than GCs with $R_{\rm PER}$ $<$ 3.5 kpc. 

\begin{figure*}
\begin{center}
\includegraphics[height=16cm,trim={0cm 1cm 4cm 2cm},clip]{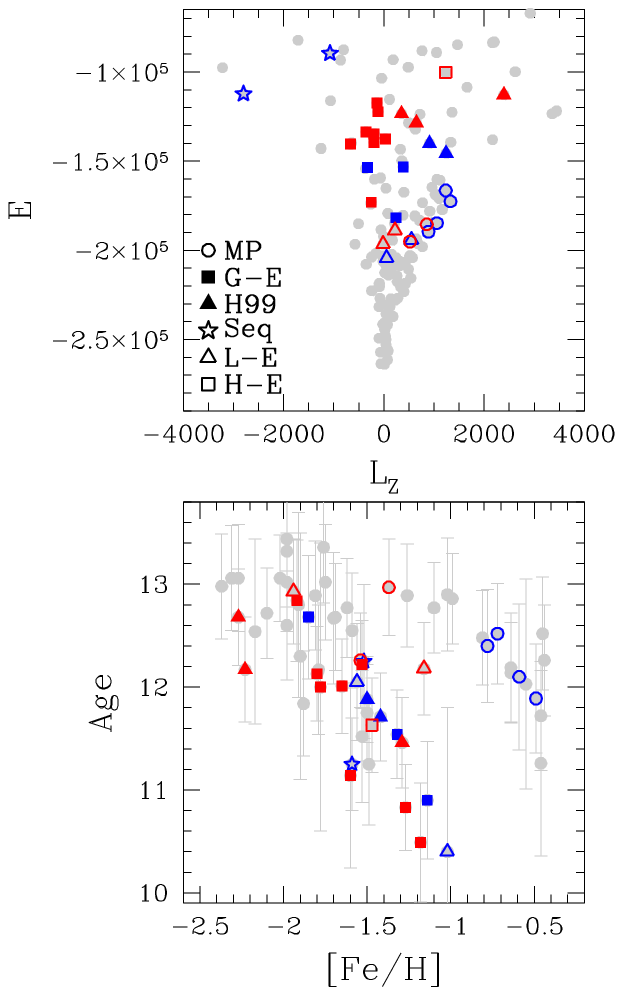}
\caption{{\it Upper panel:} E vs. ${\it L_{\rm z}}$ projections of the integral of motion space for 151 GCs (from \citet[][]{massari2019}).  Group I and Group II GCs are colored in blue and red, shape-coded according to their associations with different progenitors (open circles mark the Main Progenitor (MP), filled-squares are for {\it Gaia}-Enceladus, filled-triangles for Helmi stream (H99), asterisks for Sequoia, open-triangles for the low-energy group, and open-squares for the high energy group. {\it Lower panel:} Age-metallicity relation for Galactic GCs. Ages and metallicity are taken from \citet{kruijssen2019} and \citet[2010 version of the][catalog]{harris1996a}. Clusters are color- and shape-coded in the same way as the top panel.}
\label{fig:iom2}
\end{center}
\end{figure*}

Recently, \citet{massari2019} identified GCs with common origin by analyzing the kinematic properties provided by {\it Gaia} data in the space of integral of motion (IOM), including the energy E and the angular momentum in the z-direction, L$_{\rm z}$, which is perpendicular to the Galactic plane. We explored the link between our groups of GCs and the space of the IOM with probable progenitor galaxies of GCs provided by Massari and collaborators, together with the age-metallicity relation (AMR). Absolute ages and metallicity are from \citet{kruijssen2019, harris1996a}, respectively. The upper and bottom panels in Figure\,\ref{fig:iom2} show the IOM and AMR for the clusters in our sample, color-coded according to the two groups of GCs and shape-coded by associations with different progenitors. It is immediately clear that: 

 \begin{itemize}
  \item Clusters associated with {\it Gaia}-Enceladus (G-E) and Helmi stream (H99): Among these GCs, clusters of group II mostly have higher energy than those of group I. The only exception is NGC\,6205 with E = -1.73 $\times$ 10$^5$ and L$_{\rm z}$ = -251. 
In 100000 Monte Carlo simulations where we assumed that the simulated GCs have the same distribution as the observed clusters associated with G-E and H99 in the IOM, we find that the probability that ten or eleven out of eleven randomly extracted GCs have energy more than -1.41$\times$ 10$^5$ is 0.013. Also, the AMR of Group II appear to be slightly shifted by $\sim$ 0.2 dex toward metal-poor side from AMR of Group I GCs, but we could not confidently conclude this due to a large errors of cluster age. 

  \item Clusters associated with Main-Progenitor (MP): Due to the small sample of GCs associated with MP, it would be difficult to argue that there is a clear difference in Energy and L$_{\rm z}$  between the two groups of GCs. We noticed, however, that Group II GCs, NGC\,6218 and NGC\,6752 are definitely more metal-poor than Group I GCs as shown in the AMR.

  \item Clusters associated with Sequoia (Seq): Two clusters associated with Sequoia in our sample of GCs, NGC\,3201 and NGC\,7006, have retrograde orbits and very high energies. Unlike other GCs with high energy, these two GCs belong to Group I.
    \end{itemize}

For GCs associated with Low-Energy (L-E), no relations between the two groups of GCs and the IOM space are found, but Group II GCs seem to be older than Group I GCs. In fact, \citet{massari2019} labeled some GCs, which were not able to be associated with known merger events, L-E or H-E according to their energy level. Hence, they cannot have a common origin.

\begin{figure*}
\begin{center}

\includegraphics[height=14cm,trim={0cm 0cm 4cm 0cm},clip]{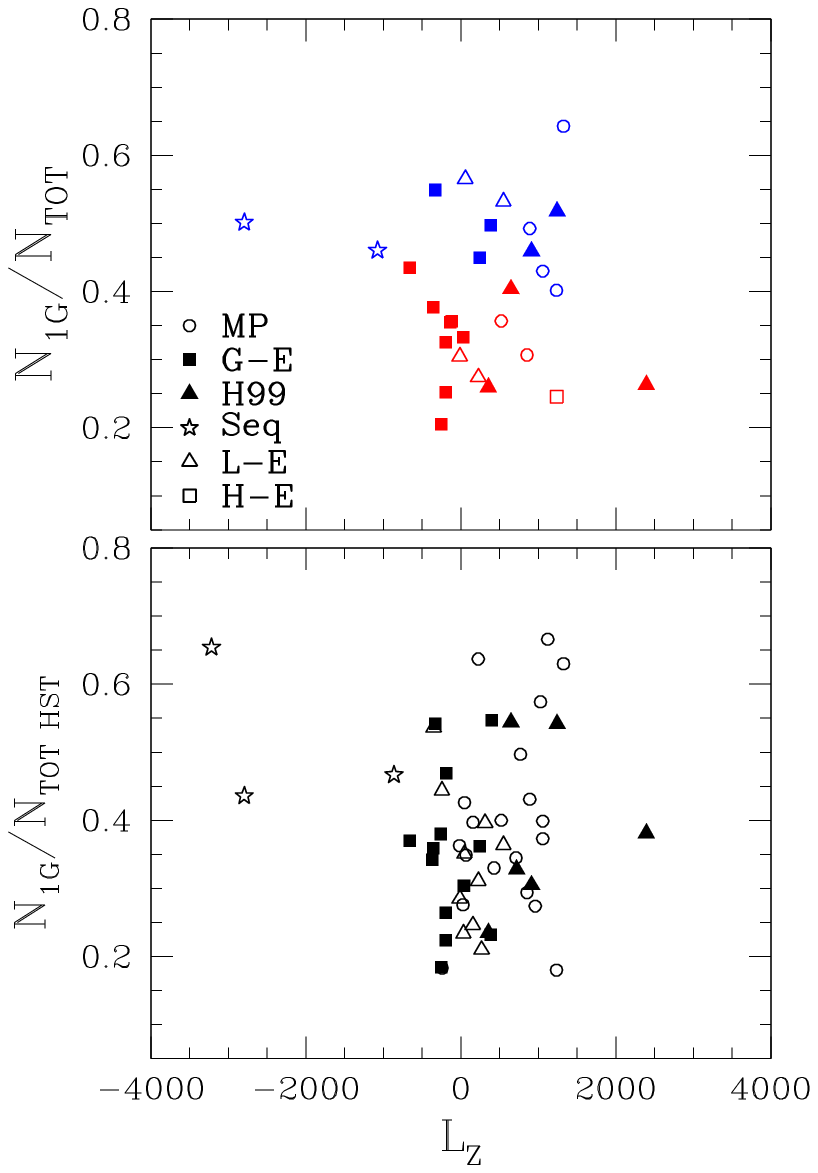}

\caption{N$_{\rm 1G}$/N$_{\rm TOT}$ vs. L$_{\rm z}$ for Galactic GCs. N$_{\rm 1G}$/N$_{\rm TOT}$ in the upper panel and the lower panel mark the results derived from ground-based and {\it HST} photometry, respectively. Group I and Group II GCs are colored in blue and red in the upper panel, respectively. Clusters are shape-coded as shown in Figure\,\ref{fig:iom2} according to associations with different progenitors.}
\label{fig:Lz}
\end{center}
\end{figure*}

The top and bottom panels in Figure\,\ref{fig:Lz} show L$_{\rm z}$ versus N$_{\rm 1G}$/N$_{\rm TOT}$ from this study and \citet{milone2017a}, respectively. We found that the fraction of 1G stars appears to mildly anticorrelate with L$_{\rm z}$ both in the Groups. This is indicated by the Spearman’s rank correlation coefficients, which are R$_{\rm s}$ = -0.10 for Group I and -0.23 for Group II. On the contrary, no correlation is found between L$_{\rm z}$ and the 1G fractions in the {\it HST} field of view.   



\section{SUMMARY AND CONCLUSIONS} \label{sec:summary}




 In this work, we have analyzed the $\Delta c_{\rm U,B,I}$ versus $\Delta_{B,I}$ pseudo two color diagrams of RGB stars for 29 GCs from \citet[see Figure 12 $\&$ 13 in][] {jang2022} to investigate multiple stellar populations over a wide field of view. The ChMs allowed us to derive the fractions of stellar populations in the wide-field spatial extent of a large sample of GCs, which have various implications on interplay between the multiple populations and the galactic environment. In particular, we explored whether the fraction of stellar populations in the analyzed field of view have been influenced by GC global parameters and the dynamic evolution of GCs in our Galaxy. The main results can be summarized as follows.

\begin{itemize}


    \item The analyzed GCs are categorized according to the radial distributions of the 2G fraction; GCs with centrally concentrated 2G stars:  NGC\,5927, NGC\,104, NGC\,2808, NGC\,1851, NGC\,6981, NGC\,5272, NGC\,6254, NGC\,3201, and NGC\,6809; GCs with spatially mixed populations: NGC\,6366, NGC\,6838, NGC\,6712, NGC\,6121, NGC\,1261, NGC\,5904, NGC\,288, NGC\,6218, NGC\,6934, NGC\,6205, NGC\,6752, NGC\,1904, NGC\,7089, NGC\,7492, NGC\,4147, NGC\,4833, NGC\,2298, NGC\,4590, and NGC\,5053. 
    NGC\,7006 with a larger fraction of 2G stars in the external region is a possible exception. However, small statistics prevent us from a firm conclusion for this cluster. Thus, we conclude that all the GCs either have a flat distribution or exhibit a more centrally concentrated 2G stars. 

    \item The resulting fractions of 1G stars in the analyzed field of view generally anticorrelate with cluster mass as already discovered in the literature, but, at the same time, they are clearly bifurcated across all mass range, allowing us to divide them into two groups of GCs. We referred to groups of clusters with higher or lower fractions as Group I and Group II, respectively.

    \item The dichotomy still exists when removing stars within the half-light radius, implying that a group of GCs with lower 1G fractions at given masses of clusters might have more efficiently lost their 1G stars in the outermost cluster regions. 

    \item We studied the link between the two groups of GCs and the categorized radial distributions of 2G fraction. We find that almost all GCs in Group II exhibit flat radial distributions of 2G stars, while the distributions tend to be flat in only less massive GCs among Group I GCs. Massive GCs in Group I have more centrally concentrated 2G stars. 

    \item Group II GCs span a wider range of $R_{\rm GC}$ and have smaller perigalactic radii $R_{\rm PER}$ $<$ 3.5 kpc, except for the most metal-poor GCs, NGC\,4590 and NGC\,5053 with the largest perigalactic radii. 

    \item We used the IOM space and probable progenitor galaxies of GCs provided by \citet{massari2019}, including AMR, to explore the link between those and the two groups of GCs defined in this work. We find that Group II GCs tend to have higher energy and to be relatively metal-poor, especially in the case of clusters associated with {\it Gaia}-Enceladus and Helmi stream. Among clusters formed in situ referred as the Main Progenitor, Group II GCs are more metal-poor than Group I GCs. 


    \item The fraction of 1G stars inferred from wide-field photometry appears to mildly anti-correlate with the z-angular momentum L$_{\rm z}$ of GCs toward the Galactic north pole both in the two group of GCs. No correlation is found with the 1G fractions in central region derived from the {\it HST} photometry.


\end{itemize}

The analysis of wide-field ground-based ChMs of RGB stars in 29 GCs provides the observational evidence of dynamic path that GCs and stellar populations in them have stepped on. Our study has first shown a clear dichotomy of GCs in the number ratio of the first population to the total number of stars analyzed at given masses of GCs. A link between the kinematic information of GCs and the two groups of GCs identified in this study implies that Group II GCs have experienced more drastic dynamic evolution, losing more stars in the outermost cluster region. Although additional work is needed to constrain the initial physical properties of clusters and kinematic path they have stepped on in detail both 
and/or in the context of different assemble history, our result provides some global view of dynamic evolution of Galactic GCs.

 \begin{table*}
\setlength{\tabcolsep}{5pt}
  \caption{Values of the fraction of 1G stars with respect to the total number of analyzed stars and the number of analyzed stars outside the half-light radius are reported in the second, third, fifth and sixth column, which are obtained from the ground-based ChMs of the 29 GCs. The fourth column provides the ratio between the median radial distance of all the analyzed stars from the cluster center (R$_{\rm median}$) and the cluster half-light radius (r$_{\rm hl}$). The second last column provides the fraction of 1G stars within the half-light radius, which is measured from {\it HST}-based ChM. We indicate Group I and II in the last column. 
  }
  \label{tab:DR}
  \begin{tabular}{ccccccccc}
    \hline
    \hline
 ID & N$_{\rm 1G}$/N$_{\rm TOT}$ & N$_{\rm stars}$ & R$_{\rm median}$/r$_{\rm hl}$ & N$_{\rm 1G}$/N$_{\rm TOT}$ ($>$ r$_{\rm hl}$)\footnote{The fraction of 1G stars in the region outside the half-light radius.} & N$_{\rm stars}$ ($>$ r$_{\rm hl}$) &  N$_{\rm 1G}$/N$_{\rm TOT}$ ($<$ r$_{\rm hl}$)\footnote{The fraction of 1G stars in the region inside the half-light radius.} & Group \\
  
   \hline

NGC  104&   0.40 $\pm$0.01 &2095 & 2.31&  0.40$\pm$0.01 & 1923 &  0.18$\pm$0.01 &1\\
NGC  288&   0.56 $\pm$0.02 & 430 & 1.24&  0.55$\pm$0.03 & 266  &  0.54$\pm$0.03 &1\\
NGC 1261&   0.39 $\pm$0.03 & 264 & 2.47&  0.38$\pm$0.03 & 251  &  0.35$\pm$0.03 &2\\
NGC 1851&   0.33 $\pm$0.03 & 353 & 4.88&  0.33$\pm$0.03 & 353  &  0.26$\pm$0.03 &2\\
NGC 1904&   0.33 $\pm$0.03 & 319 & 2.97&  0.33$\pm$0.03 & 313  &         -       &2\\
NGC 2298&   0.48 $\pm$0.04 & 164 & 1.46&  0.44$\pm$0.04 & 116  &  0.41$\pm$0.07 &2\\
NGC 2808&   0.50 $\pm$0.02 & 942 & 4.16&  0.50$\pm$0.02 & 942  &  0.22$\pm$0.02 &1\\
NGC 3201&   0.56 $\pm$0.04 & 143 & 1.04&  0.50$\pm$0.05 &  75  &  0.44$\pm$0.04 &1\\
NGC 4147&   0.36 $\pm$0.04 &  92 & 2.08&  0.36$\pm$0.05 &  78  &         -       &2\\
NGC 4590&   0.33 $\pm$0.04 & 143 & 1.22&  0.26$\pm$0.05 &  87  &  0.39$\pm$0.04 &2\\
NGC 4833&   0.53 $\pm$0.04 & 194 & 1.05&  0.45$\pm$0.05 & 102  &  0.36$\pm$0.03 &1\\
NGC 5053&   0.42 $\pm$0.03 & 236 & 0.82&  0.40$\pm$0.05 &  95  &  0.54$\pm$0.06 &2\\
NGC 5272&   0.46 $\pm$0.03 & 292 & 1.42&  0.46$\pm$0.03 & 202  &  0.31$\pm$0.01 &1\\
NGC 5904&   0.29 $\pm$0.03 & 396 & 2.08&  0.26$\pm$0.03 & 345  &  0.24$\pm$0.02 &2\\
NGC 5927&   0.43 $\pm$0.02 & 662 & 2.30&  0.43$\pm$0.02 & 625  &  0.38$\pm$0.04 &1\\
NGC 6121&   0.31 $\pm$0.03 & 289 & 0.68&  0.30$\pm$0.05 &  81  &  0.29$\pm$0.04 &2\\
NGC 6205&   0.21 $\pm$0.02 & 526 & 2.25&  0.21$\pm$0.02 & 494  &  0.19$\pm$0.01 &2\\
NGC 6218&   0.38 $\pm$0.03 & 343 & 1.48&  0.35$\pm$0.03 & 240  &  0.40$\pm$0.03 &2\\
NGC 6254&   0.51 $\pm$0.02 & 624 & 1.64&  0.53$\pm$0.02 & 474  &  0.36$\pm$0.03 &1\\
NGC 6366&   0.56 $\pm$0.05 & 105 & 0.93&  0.49$\pm$0.06 &  47  &  0.43$\pm$0.05 &1\\
NGC 6712&   0.61 $\pm$0.03 & 304 & 1.28&  0.57$\pm$0.04 & 194  &         -      &1\\
NGC 6752&   0.33 $\pm$0.02 & 543 & 1.96&  0.31$\pm$0.02 & 452  &  0.29$\pm$0.03 &2\\
NGC 6809&   0.24 $\pm$0.02 & 347 & 1.06&  0.27$\pm$0.03 & 187  &  0.31$\pm$0.03 &2\\
NGC 6838&   0.66 $\pm$0.03 & 224 & 0.98&  0.64$\pm$0.05 & 106  &  0.62$\pm$0.09 &1\\
NGC 6934&   0.26 $\pm$0.02 & 284 & 2.06&  0.25$\pm$0.03 & 264  &  0.30$\pm$0.03 &2\\
NGC 6981&   0.49 $\pm$0.04 & 194 & 1.58&  0.52$\pm$0.04 & 144  &  0.56$\pm$0.05 &1\\
NGC 7006&   0.52 $\pm$0.07 &  55 & 1.57&  0.46$\pm$0.09 &  46  &         -       &1\\
NGC 7089&   0.28 $\pm$0.02 & 451 & 2.52&  0.25$\pm$0.02 & 433  &  0.22$\pm$0.02 &2\\
NGC 7492&   0.36 $\pm$0.04 & 117 & 0.99&  0.36$\pm$0.07 &  55  &        -        &2\\

 \hline
 \hline
\end{tabular}
 \end{table*}


\begin{acknowledgments}
This work has received funding from the NRF of Korea (2022R1A2C3002992, 2022R1A6A1A03053472). APM, GC, and AFM acknowledge the support received from the European Union’s Horizon 2020 research and innovation programme under the Marie Sk\l{}odowska-Curie Grant Agreement No. 101034319 and from INAF Research GTO-Grant Normal RSN2-1.05.12.05.10 – Understanding
the formation of globular clusters with their multiple stellar generations (ref. AFM) of the ‘Bando INAF per il Finanziamento della Ricerca Fondamentale 2022’. 
EPL acknowledges support from the ``Science \& Technology Champion Project” (202005AB160002) and from the ``Top Team Project” (202305AT350002), all funded by the ``Yunnan Revitalization Talent Support Program”.  
\end{acknowledgments}

%





\bibliography{ref}{}

\begin{thebibliography}{}
\expandafter\ifx\csname natexlab\endcsname\relax\def\natexlab#1{#1}\fi
\providecommand{\url}[1]{\href{#1}{#1}}
\providecommand{\dodoi}[1]{doi:~\href{http://doi.org/#1}{\nolinkurl{#1}}}
\providecommand{\doeprint}[1]{\href{http://ascl.net/#1}{\nolinkurl{http://ascl.net/#1}}}
\providecommand{\doarXiv}[1]{\href{https://arxiv.org/abs/#1}{\nolinkurl{https://arxiv.org/abs/#1}}}

\bibitem[{{Alessandrini} {et~al.}(2016){Alessandrini}, {Lanzoni}, {Ferraro}, {Miocchi}, \& {Vesperini}}]{Alessandrini2016}
{Alessandrini}, E., {Lanzoni}, B., {Ferraro}, F.~R., {Miocchi}, P., \& {Vesperini}, E. 2016, \apj, 833, 252, \dodoi{10.3847/1538-4357/833/2/252}

\bibitem[{{Bastian} {et~al.}(2018){Bastian}, {Kamann}, {Cabrera-Ziri}, {Georgy}, {Ekstr{\"o}m}, {Charbonnel}, {de Juan Ovelar}, \& {Usher}}]{bastian2018a}
{Bastian}, N., {Kamann}, S., {Cabrera-Ziri}, I., {et~al.} 2018, \mnras, 480, 3739, \dodoi{10.1093/mnras/sty2100}

\bibitem[{{Bastian} {et~al.}(2013){Bastian}, {Lamers}, {de Mink}, {Longmore}, {Goodwin}, \& {Gieles}}]{bastian2013a}
{Bastian}, N., {Lamers}, H.~J.~G.~L.~M., {de Mink}, S.~E., {et~al.} 2013, \mnras, 436, 2398, \dodoi{10.1093/mnras/stt1745}

\bibitem[{{Baumgardt}(2017)}]{Baumgardt2017}
{Baumgardt}, H. 2017, \mnras, 464, 2174, \dodoi{10.1093/mnras/stw2488}

\bibitem[{{Baumgardt} \& {Hilker}(2018)}]{baumgardt2018a}
{Baumgardt}, H., \& {Hilker}, M. 2018, \mnras, 478, 1520, \dodoi{10.1093/mnras/sty1057}

\bibitem[{{Baumgardt} {et~al.}(2019){Baumgardt}, {Hilker}, {Sollima}, \& {Bellini}}]{baumgardt2019}
{Baumgardt}, H., {Hilker}, M., {Sollima}, A., \& {Bellini}, A. 2019, \mnras, 482, 5138, \dodoi{10.1093/mnras/sty2997}

\bibitem[{{Carretta} {et~al.}(2010){Carretta}, {Bragaglia}, {D'Orazi}, {Lucatello}, \& {Gratton}}]{carretta2010}
{Carretta}, E., {Bragaglia}, A., {D'Orazi}, V., {Lucatello}, S., \& {Gratton}, R.~G. 2010, \aap, 519, A71, \dodoi{10.1051/0004-6361/201014996}

\bibitem[{{Carretta} {et~al.}(2009){Carretta}, {Bragaglia}, {Gratton}, {Lucatello}, {Catanzaro}, {Leone}, {Bellazzini}, {Claudi}, {D'Orazi}, {Momany}, {Ortolani}, {Pancino}, {Piotto}, {Recio-Blanco}, \& {Sabbi}}]{carretta2009a}
{Carretta}, E., {Bragaglia}, A., {Gratton}, R.~G., {et~al.} 2009, \aap, 505, 117, \dodoi{10.1051/0004-6361/200912096}

\bibitem[{{Conroy}(2012)}]{conroy2012}
{Conroy}, C. 2012, \apj, 758, 21, \dodoi{10.1088/0004-637X/758/1/21}

\bibitem[{{Cordoni} {et~al.}(2020){Cordoni}, {Milone}, {Marino}, {Da Costa}, {Dondoglio}, {Jerjen}, {Lagioia}, {Mastrobuono-Battisti}, {Norris}, {Tailo}, \& {Yong}}]{cordoni2020b}
{Cordoni}, G., {Milone}, A.~P., {Marino}, A.~F., {et~al.} 2020, \apj, 898, 147, \dodoi{10.3847/1538-4357/aba04b}

\bibitem[{{Cordoni} {et~al.}(2022){Cordoni}, {Milone}, {Marino}, {Cignoni}, {Lagioia}, {Tailo}, {Carlos}, {Dondoglio}, {Jang}, {Mohandasan}, \& {Legnardi}}]{cordoni2022}
---. 2022, Nature Communications, 13, 4325, \dodoi{10.1038/s41467-022-31977-y}

\bibitem[{{Cordoni} {et~al.}(2023){Cordoni}, {Marino}, {Milone}, {Dondoglio}, {Lagioia}, {Legnardi}, {Mohandasan}, {Jang}, \& {Ziliotto}}]{cordoni2023}
{Cordoni}, G., {Marino}, A.~F., {Milone}, A.~P., {et~al.} 2023, \aap, 678, A155, \dodoi{10.1051/0004-6361/202347189}

\bibitem[{{Cordoni} {et~al.}(2024){Cordoni}, {Casagrande}, {Milone}, {Dondoglio}, {Mastrobuono-Battisti}, {Jang}, {Marino}, {Lagioia}, {Vittoria Legnardi}, {Ziliotto}, {Muratore}, {Mehta}, {Lacchin}, \& {Tailo}}]{cordoni2024}
{Cordoni}, G., {Casagrande}, L., {Milone}, A., {et~al.} 2024, arXiv e-prints, arXiv:2409.02330, \dodoi{10.48550/arXiv.2409.02330}

\bibitem[{{Dalessandro} {et~al.}(2019){Dalessandro}, {Cadelano}, {Vesperini}, {Martocchia}, {Ferraro}, {Lanzoni}, {Bastian}, {Hong}, \& {Sanna}}]{dalessandro2019a}
{Dalessandro}, E., {Cadelano}, M., {Vesperini}, E., {et~al.} 2019, \apjl, 884, L24, \dodoi{10.3847/2041-8213/ab45f7}

\bibitem[{{D'Antona} {et~al.}(2016){D'Antona}, {Vesperini}, {D'Ercole}, {Ventura}, {Milone}, {Marino}, \& {Tailo}}]{dantona2016a}
{D'Antona}, F., {Vesperini}, E., {D'Ercole}, A., {et~al.} 2016, \mnras, 458, 2122, \dodoi{10.1093/mnras/stw387}

\bibitem[{{de Mink} {et~al.}(2009){de Mink}, {Pols}, {Langer}, \& {Izzard}}]{demink2009a}
{de Mink}, S.~E., {Pols}, O.~R., {Langer}, N., \& {Izzard}, R.~G. 2009, \aap, 507, L1, \dodoi{10.1051/0004-6361/200913205}

\bibitem[{{Decressin} {et~al.}(2007){Decressin}, {Meynet}, {Charbonnel}, {Prantzos}, \& {Ekstr{\"o}m}}]{decressin2007a}
{Decressin}, T., {Meynet}, G., {Charbonnel}, C., {Prantzos}, N., \& {Ekstr{\"o}m}, S. 2007, \aap, 464, 1029, \dodoi{10.1051/0004-6361:20066013}

\bibitem[{{Denissenkov} \& {Hartwick}(2014)}]{denissenkov2014a}
{Denissenkov}, P.~A., \& {Hartwick}, F.~D.~A. 2014, \mnras, 437, L21, \dodoi{10.1093/mnrasl/slt133}

\bibitem[{{D'Ercole} {et~al.}(2008){D'Ercole}, {Vesperini}, {D'Antona}, {McMillan}, \& {Recchi}}]{dercole2008a}
{D'Ercole}, A., {Vesperini}, E., {D'Antona}, F., {McMillan}, S. L.~W., \& {Recchi}, S. 2008, \mnras, 391, 825, \dodoi{10.1111/j.1365-2966.2008.13915.x}

\bibitem[{{Dondoglio} {et~al.}(2021){Dondoglio}, {Milone}, {Lagioia}, {Marino}, {Tailo}, {Cordoni}, {Jang}, \& {Carlos}}]{dondoglio2021a}
{Dondoglio}, E., {Milone}, A.~P., {Lagioia}, E.~P., {et~al.} 2021, \apj, 906, 76, \dodoi{10.3847/1538-4357/abc882}

\bibitem[{{Dondoglio} {et~al.}(2023){Dondoglio}, {Milone}, {Marino}, {D'Antona}, {Cordoni}, {Legnardi}, {Lagioia}, {Jang}, {Ziliotto}, {Carlos}, {Dell'Agli}, {Karakas}, {Mohandasan}, {Osborn}, {Tailo}, \& {Ventura}}]{dondoglio2023}
{Dondoglio}, E., {Milone}, A.~P., {Marino}, A.~F., {et~al.} 2023, \mnras, 526, 2960, \dodoi{10.1093/mnras/stad2950}

\bibitem[{{Gaia Collaboration} {et~al.}(2018){Gaia Collaboration}, {Brown}, {Vallenari}, {Prusti}, {de Bruijne}, {Babusiaux}, {Bailer-Jones}, {Biermann}, {Evans}, {Eyer}, {Jansen}, {Jordi}, {Klioner}, {Lammers}, {Lindegren}, {Luri}, {Mignard}, {Panem}, {Pourbaix}, {Randich}, {Sartoretti}, {Siddiqui}, {Soubiran}, {van Leeuwen}, {Walton}, {Arenou}, {Bastian}, {Cropper}, {Drimmel}, {Katz}, {Lattanzi}, {Bakker}, {Cacciari}, {Casta{\~n}eda}, {Chaoul}, {Cheek}, {De Angeli}, {Fabricius}, {Guerra}, {Holl}, {Masana}, {Messineo}, {Mowlavi}, {Nienartowicz}, {Panuzzo}, {Portell}, {Riello}, {Seabroke}, {Tanga}, {Th{\'e}venin}, {Gracia-Abril}, {Comoretto}, {Garcia-Reinaldos}, {Teyssier}, {Altmann}, {Andrae}, {Audard}, {Bellas-Velidis}, {Benson}, {Berthier}, {Blomme}, {Burgess}, {Busso}, {Carry}, {Cellino}, {Clementini}, {Clotet}, {Creevey}, {Davidson}, {De Ridder}, {Delchambre}, {Dell'Oro}, {Ducourant}, {Fern{\'a}ndez-Hern{\'a}ndez}, {Fouesneau}, {Fr{\'e}mat}, {Galluccio}, {Garc{\'\i}a-Torres},
  {Gonz{\'a}lez-N{\'u}{\~n}ez}, {Gonz{\'a}lez-Vidal}, {Gosset}, {Guy}, {Halbwachs}, {Hambly}, {Harrison}, {Hern{\'a}ndez}, {Hestroffer}, {Hodgkin}, {Hutton}, {Jasniewicz}, {Jean-Antoine-Piccolo}, {Jordan}, {Korn}, {Krone-Martins}, {Lanzafame}, {Lebzelter}, {L{\"o}ffler}, {Manteiga}, {Marrese}, {Mart{\'\i}n-Fleitas}, {Moitinho}, {Mora}, {Muinonen}, {Osinde}, {Pancino}, {Pauwels}, {Petit}, {Recio-Blanco}, {Richards}, {Rimoldini}, {Robin}, {Sarro}, {Siopis}, {Smith}, {Sozzetti}, {S{\"u}veges}, {Torra}, {van Reeven}, {Abbas}, {Abreu Aramburu}, {Accart}, {Aerts}, {Altavilla}, {{\'A}lvarez}, {Alvarez}, {Alves}, {Anderson}, {Andrei}, {Anglada Varela}, {Antiche}, {Antoja}, {Arcay}, {Astraatmadja}, {Bach}, {Baker}, {Balaguer-N{\'u}{\~n}ez}, {Balm}, {Barache}, {Barata}, {Barbato}, {Barblan}, {Barklem}, {Barrado}, {Barros}, {Barstow}, {Bartholom{\'e} Mu{\~n}oz}, {Bassilana}, {Becciani}, {Bellazzini}, {Berihuete}, {Bertone}, {Bianchi}, {Bienaym{\'e}}, {Blanco-Cuaresma}, {Boch}, {Boeche}, {Bombrun}, {Borrachero},
  {Bossini}, {Bouquillon}, {Bourda}, {Bragaglia}, {Bramante}, {Breddels}, {Bressan}, {Brouillet}, {Br{\"u}semeister}, {Brugaletta}, {Bucciarelli}, {Burlacu}, {Busonero}, {Butkevich}, {Buzzi}, {Caffau}, {Cancelliere}, {Cannizzaro}, {Cantat-Gaudin}, {Carballo}, {Carlucci}, {Carrasco}, {Casamiquela}, {Castellani}, {Castro-Ginard}, {Charlot}, {Chemin}, {Chiavassa}, {Cocozza}, {Costigan}, {Cowell}, {Crifo}, {Crosta}, {Crowley}, {Cuypers}, {Dafonte}, {Damerdji}, {Dapergolas}, {David}, {David}, {de Laverny}, {De Luise}, {De March}, {de Martino}, {de Souza}, {de Torres}, {Debosscher}, {del Pozo}, {Delbo}, {Delgado}, {Delgado}, {Di Matteo}, {Diakite}, {Diener}, {Distefano}, {Dolding}, {Drazinos}, {Dur{\'a}n}, {Edvardsson}, {Enke}, {Eriksson}, {Esquej}, {Eynard Bontemps}, {Fabre}, {Fabrizio}, {Faigler}, {Falc{\~a}o}, {Farr{\`a}s Casas}, {Federici}, {Fedorets}, {Fernique}, {Figueras}, {Filippi}, {Findeisen}, {Fonti}, {Fraile}, {Fraser}, {Fr{\'e}zouls}, {Gai}, {Galleti}, {Garabato}, {Garc{\'\i}a-Sedano}, {Garofalo},
  {Garralda}, {Gavel}, {Gavras}, {Gerssen}, {Geyer}, {Giacobbe}, {Gilmore}, {Girona}, {Giuffrida}, {Glass}, {Gomes}, {Granvik}, {Gueguen}, {Guerrier}, {Guiraud}, {Guti{\'e}rrez-S{\'a}nchez}, {Haigron}, {Hatzidimitriou}, {Hauser}, {Haywood}, {Heiter}, {Helmi}, {Heu}, {Hilger}, {Hobbs}, {Hofmann}, {Holland}, {Huckle}, {Hypki}, {Icardi}, {Jan{\ss}en}, {Jevardat de Fombelle}, {Jonker}, {Juh{\'a}sz}, {Julbe}, {Karampelas}, {Kewley}, {Klar}, {Kochoska}, {Kohley}, {Kolenberg}, {Kontizas}, {Kontizas}, {Koposov}, {Kordopatis}, {Kostrzewa-Rutkowska}, {Koubsky}, {Lambert}, {Lanza}, {Lasne}, {Lavigne}, {Le Fustec}, {Le Poncin-Lafitte}, {Lebreton}, {Leccia}, {Leclerc}, {Lecoeur-Taibi}, {Lenhardt}, {Leroux}, {Liao}, {Licata}, {Lindstr{\o}m}, {Lister}, {Livanou}, {Lobel}, {L{\'o}pez}, {Managau}, {Mann}, {Mantelet}, {Marchal}, {Marchant}, {Marconi}, {Marinoni}, {Marschalk{\'o}}, {Marshall}, {Martino}, {Marton}, {Mary}, {Massari}, {Matijevi{\v{c}}}, {Mazeh}, {McMillan}, {Messina}, {Michalik}, {Millar}, {Molina}, {Molinaro},
  {Moln{\'a}r}, {Montegriffo}, {Mor}, {Morbidelli}, {Morel}, {Morris}, {Mulone}, {Muraveva}, {Musella}, {Nelemans}, {Nicastro}, {Noval}, {O'Mullane}, {Ord{\'e}novic}, {Ord{\'o}{\~n}ez-Blanco}, {Osborne}, {Pagani}, {Pagano}, {Pailler}, {Palacin}, {Palaversa}, {Panahi}, {Pawlak}, {Piersimoni}, {Pineau}, {Plachy}, {Plum}, {Poggio}, {Poujoulet}, {Pr{\v{s}}a}, {Pulone}, {Racero}, {Ragaini}, {Rambaux}, {Ramos-Lerate}, {Regibo}, {Reyl{\'e}}, {Riclet}, {Ripepi}, {Riva}, {Rivard}, {Rixon}, {Roegiers}, {Roelens}, {Romero-G{\'o}mez}, {Rowell}, {Royer}, {Ruiz-Dern}, {Sadowski}, {Sagrist{\`a} Sell{\'e}s}, {Sahlmann}, {Salgado}, {Salguero}, {Sanna}, {Santana-Ros}, {Sarasso}, {Savietto}, {Schultheis}, {Sciacca}, {Segol}, {Segovia}, {S{\'e}gransan}, {Shih}, {Siltala}, {Silva}, {Smart}, {Smith}, {Solano}, {Solitro}, {Sordo}, {Soria Nieto}, {Souchay}, {Spagna}, {Spoto}, {Stampa}, {Steele}, {Steidelm{\"u}ller}, {Stephenson}, {Stoev}, {Suess}, {Surdej}, {Szabados}, {Szegedi-Elek}, {Tapiador}, {Taris}, {Tauran}, {Taylor},
  {Teixeira}, {Terrett}, {Teyssandier}, {Thuillot}, {Titarenko}, {Torra Clotet}, {Turon}, {Ulla}, {Utrilla}, {Uzzi}, {Vaillant}, {Valentini}, {Valette}, {van Elteren}, {Van Hemelryck}, {van Leeuwen}, {Vaschetto}, {Vecchiato}, {Veljanoski}, {Viala}, {Vicente}, {Vogt}, {von Essen}, {Voss}, {Votruba}, {Voutsinas}, {Walmsley}, {Weiler}, {Wertz}, {Wevers}, {Wyrzykowski}, {Yoldas}, {{\v{Z}}erjal}, {Ziaeepour}, {Zorec}, {Zschocke}, {Zucker}, {Zurbach}, \& {Zwitter}}]{gaiaDR2}
{Gaia Collaboration}, {Brown}, A.~G.~A., {Vallenari}, A., {et~al.} 2018, \aap, 616, A1, \dodoi{10.1051/0004-6361/201833051}

\bibitem[{{Gaia Collaboration} {et~al.}(2021){Gaia Collaboration}, {Brown}, {Vallenari}, {Prusti}, {de Bruijne}, {Babusiaux}, {Biermann}, {Creevey}, {Evans}, {Eyer}, {Hutton}, {Jansen}, {Jordi}, {Klioner}, {Lammers}, {Lindegren}, {Luri}, {Mignard}, {Panem}, {Pourbaix}, {Randich}, {Sartoretti}, {Soubiran}, {Walton}, {Arenou}, {Bailer-Jones}, {Bastian}, {Cropper}, {Drimmel}, {Katz}, {Lattanzi}, {van Leeuwen}, {Bakker}, {Cacciari}, {Casta{\~n}eda}, {De Angeli}, {Ducourant}, {Fabricius}, {Fouesneau}, {Fr{\'e}mat}, {Guerra}, {Guerrier}, {Guiraud}, {Jean-Antoine Piccolo}, {Masana}, {Messineo}, {Mowlavi}, {Nicolas}, {Nienartowicz}, {Pailler}, {Panuzzo}, {Riclet}, {Roux}, {Seabroke}, {Sordo}, {Tanga}, {Th{\'e}venin}, {Gracia-Abril}, {Portell}, {Teyssier}, {Altmann}, {Andrae}, {Bellas-Velidis}, {Benson}, {Berthier}, {Blomme}, {Brugaletta}, {Burgess}, {Busso}, {Carry}, {Cellino}, {Cheek}, {Clementini}, {Damerdji}, {Davidson}, {Delchambre}, {Dell'Oro}, {Fern{\'a}ndez-Hern{\'a}ndez}, {Galluccio}, {Garc{\'\i}a-Lario},
  {Garcia-Reinaldos}, {Gonz{\'a}lez-N{\'u}{\~n}ez}, {Gosset}, {Haigron}, {Halbwachs}, {Hambly}, {Harrison}, {Hatzidimitriou}, {Heiter}, {Hern{\'a}ndez}, {Hestroffer}, {Hodgkin}, {Holl}, {Jan{\ss}en}, {Jevardat de Fombelle}, {Jordan}, {Krone-Martins}, {Lanzafame}, {L{\"o}ffler}, {Lorca}, {Manteiga}, {Marchal}, {Marrese}, {Moitinho}, {Mora}, {Muinonen}, {Osborne}, {Pancino}, {Pauwels}, {Petit}, {Recio-Blanco}, {Richards}, {Riello}, {Rimoldini}, {Robin}, {Roegiers}, {Rybizki}, {Sarro}, {Siopis}, {Smith}, {Sozzetti}, {Ulla}, {Utrilla}, {van Leeuwen}, {van Reeven}, {Abbas}, {Abreu Aramburu}, {Accart}, {Aerts}, {Aguado}, {Ajaj}, {Altavilla}, {{\'A}lvarez}, {{\'A}lvarez Cid-Fuentes}, {Alves}, {Anderson}, {Anglada Varela}, {Antoja}, {Audard}, {Baines}, {Baker}, {Balaguer-N{\'u}{\~n}ez}, {Balbinot}, {Balog}, {Barache}, {Barbato}, {Barros}, {Barstow}, {Bartolom{\'e}}, {Bassilana}, {Bauchet}, {Baudesson-Stella}, {Becciani}, {Bellazzini}, {Bernet}, {Bertone}, {Bianchi}, {Blanco-Cuaresma}, {Boch}, {Bombrun}, {Bossini},
  {Bouquillon}, {Bragaglia}, {Bramante}, {Breedt}, {Bressan}, {Brouillet}, {Bucciarelli}, {Burlacu}, {Busonero}, {Butkevich}, {Buzzi}, {Caffau}, {Cancelliere}, {C{\'a}novas}, {Cantat-Gaudin}, {Carballo}, {Carlucci}, {Carnerero}, {Carrasco}, {Casamiquela}, {Castellani}, {Castro-Ginard}, {Castro Sampol}, {Chaoul}, {Charlot}, {Chemin}, {Chiavassa}, {Cioni}, {Comoretto}, {Cooper}, {Cornez}, {Cowell}, {Crifo}, {Crosta}, {Crowley}, {Dafonte}, {Dapergolas}, {David}, {David}, {de Laverny}, {De Luise}, {De March}, {De Ridder}, {de Souza}, {de Teodoro}, {de Torres}, {del Peloso}, {del Pozo}, {Delbo}, {Delgado}, {Delgado}, {Delisle}, {Di Matteo}, {Diakite}, {Diener}, {Distefano}, {Dolding}, {Eappachen}, {Edvardsson}, {Enke}, {Esquej}, {Fabre}, {Fabrizio}, {Faigler}, {Fedorets}, {Fernique}, {Fienga}, {Figueras}, {Fouron}, {Fragkoudi}, {Fraile}, {Franke}, {Gai}, {Garabato}, {Garcia-Gutierrez}, {Garc{\'\i}a-Torres}, {Garofalo}, {Gavras}, {Gerlach}, {Geyer}, {Giacobbe}, {Gilmore}, {Girona}, {Giuffrida}, {Gomel}, {Gomez},
  {Gonzalez-Santamaria}, {Gonz{\'a}lez-Vidal}, {Granvik}, {Guti{\'e}rrez-S{\'a}nchez}, {Guy}, {Hauser}, {Haywood}, {Helmi}, {Hidalgo}, {Hilger}, {H{\l}adczuk}, {Hobbs}, {Holland}, {Huckle}, {Jasniewicz}, {Jonker}, {Juaristi Campillo}, {Julbe}, {Karbevska}, {Kervella}, {Khanna}, {Kochoska}, {Kontizas}, {Kordopatis}, {Korn}, {Kostrzewa-Rutkowska}, {Kruszy{\'n}ska}, {Lambert}, {Lanza}, {Lasne}, {Le Campion}, {Le Fustec}, {Lebreton}, {Lebzelter}, {Leccia}, {Leclerc}, {Lecoeur-Taibi}, {Liao}, {Licata}, {Lindstr{\o}m}, {Lister}, {Livanou}, {Lobel}, {Madrero Pardo}, {Managau}, {Mann}, {Marchant}, {Marconi}, {Marcos Santos}, {Marinoni}, {Marocco}, {Marshall}, {Martin Polo}, {Mart{\'\i}n-Fleitas}, {Masip}, {Massari}, {Mastrobuono-Battisti}, {Mazeh}, {McMillan}, {Messina}, {Michalik}, {Millar}, {Mints}, {Molina}, {Molinaro}, {Moln{\'a}r}, {Montegriffo}, {Mor}, {Morbidelli}, {Morel}, {Morris}, {Mulone}, {Munoz}, {Muraveva}, {Murphy}, {Musella}, {Noval}, {Ord{\'e}novic}, {Orr{\`u}}, {Osinde}, {Pagani}, {Pagano},
  {Palaversa}, {Palicio}, {Panahi}, {Pawlak}, {Pe{\~n}alosa Esteller}, {Penttil{\"a}}, {Piersimoni}, {Pineau}, {Plachy}, {Plum}, {Poggio}, {Poretti}, {Poujoulet}, {Pr{\v{s}}a}, {Pulone}, {Racero}, {Ragaini}, {Rainer}, {Raiteri}, {Rambaux}, {Ramos}, {Ramos-Lerate}, {Re Fiorentin}, {Regibo}, {Reyl{\'e}}, {Ripepi}, {Riva}, {Rixon}, {Robichon}, {Robin}, {Roelens}, {Rohrbasser}, {Romero-G{\'o}mez}, {Rowell}, {Royer}, {Rybicki}, {Sadowski}, {Sagrist{\`a} Sell{\'e}s}, {Sahlmann}, {Salgado}, {Salguero}, {Samaras}, {Sanchez Gimenez}, {Sanna}, {Santove{\~n}a}, {Sarasso}, {Schultheis}, {Sciacca}, {Segol}, {Segovia}, {S{\'e}gransan}, {Semeux}, {Shahaf}, {Siddiqui}, {Siebert}, {Siltala}, {Slezak}, {Smart}, {Solano}, {Solitro}, {Souami}, {Souchay}, {Spagna}, {Spoto}, {Steele}, {Steidelm{\"u}ller}, {Stephenson}, {S{\"u}veges}, {Szabados}, {Szegedi-Elek}, {Taris}, {Tauran}, {Taylor}, {Teixeira}, {Thuillot}, {Tonello}, {Torra}, {Torra}, {Turon}, {Unger}, {Vaillant}, {van Dillen}, {Vanel}, {Vecchiato}, {Viala}, {Vicente},
  {Voutsinas}, {Weiler}, {Wevers}, {Wyrzykowski}, {Yoldas}, {Yvard}, {Zhao}, {Zorec}, {Zucker}, {Zurbach}, \& {Zwitter}}]{gaia2021a}
---. 2021, \aap, 649, A1, \dodoi{10.1051/0004-6361/202039657}

\bibitem[{{Gieles} {et~al.}(2018){Gieles}, {Charbonnel}, {Krause}, {H{\'e}nault-Brunet}, {Agertz}, {Lamers}, {Bastian}, {Gualand ris}, {Zocchi}, \& {Petts}}]{gieles2018a}
{Gieles}, M., {Charbonnel}, C., {Krause}, M. G.~H., {et~al.} 2018, \mnras, 478, 2461, \dodoi{10.1093/mnras/sty1059}

\bibitem[{{Gratton} {et~al.}(2019){Gratton}, {Bragaglia}, {Carretta}, {D'Orazi}, {Lucatello}, \& {Sollima}}]{gratton2019a}
{Gratton}, R., {Bragaglia}, A., {Carretta}, E., {et~al.} 2019, \aapr, 27, 8, \dodoi{10.1007/s00159-019-0119-3}

\bibitem[{{Gratton} {et~al.}(2004){Gratton}, {Sneden}, \& {Carretta}}]{gratton2004}
{Gratton}, R., {Sneden}, C., \& {Carretta}, E. 2004, \araa, 42, 385, \dodoi{10.1146/annurev.astro.42.053102.133945}

\bibitem[{{Grundahl} {et~al.}(1998){Grundahl}, {VandenBerg}, \& {Andersen}}]{grundahl1998a}
{Grundahl}, F., {VandenBerg}, D.~A., \& {Andersen}, M.~I. 1998, \apjl, 500, L179, \dodoi{10.1086/311419}

\bibitem[{{Harris}(1996)}]{harris1996a}
{Harris}, W.~E. 1996, \aj, 112, 1487, \dodoi{10.1086/118116}

\bibitem[{{Hartmann} {et~al.}(2022){Hartmann}, {Bonatto}, {Chies-Santos}, {Alonso-Garc{\'\i}a}, {Bastian}, {Overzier}, {Schoenell}, {Coelho}, {Branco}, {Kanaan}, {Mendes de Oliveira}, \& {Ribeiro}}]{hartmann2022a}
{Hartmann}, E.~A., {Bonatto}, C.~J., {Chies-Santos}, A.~L., {et~al.} 2022, \mnras, 515, 4191, \dodoi{10.1093/mnras/stac1411}

\bibitem[{{Jang} \& {Lee}(2015)}]{jang2015}
{Jang}, S., \& {Lee}, Y.-W. 2015, \apjs, 218, 31, \dodoi{10.1088/0067-0049/218/2/31}

\bibitem[{{Jang} {et~al.}(2014){Jang}, {Lee}, {Joo}, \& {Na}}]{jang2014a}
{Jang}, S., {Lee}, Y.~W., {Joo}, S.~J., \& {Na}, C. 2014, \mnras, 443, L15, \dodoi{10.1093/mnrasl/slu064}

\bibitem[{{Jang} {et~al.}(2021){Jang}, {Milone}, {Lagioia}, {Tailo}, {Carlos}, {Dondoglio}, {Martorano}, {Mohandasan}, {Marino}, {Cordoni}, \& {Lee}}]{jang2021a}
{Jang}, S., {Milone}, A.~P., {Lagioia}, E.~P., {et~al.} 2021, \apj, 920, 129, \dodoi{10.3847/1538-4357/ac1861}

\bibitem[{{Jang} {et~al.}(2022){Jang}, {Milone}, {Legnardi}, {Marino}, {Mastrobuono-Battisti}, {Dondoglio}, {Lagioia}, {Casagrande}, {Carlos}, {Mohandasan}, {Cordoni}, {Bortolan}, \& {Lee}}]{jang2022}
{Jang}, S., {Milone}, A.~P., {Legnardi}, M.~V., {et~al.} 2022, \mnras, 517, 5687, \dodoi{10.1093/mnras/stac3086}

\bibitem[{{Kamann} {et~al.}(2020){Kamann}, {Giesers}, {Bastian}, {Brinchmann}, {Dreizler}, {G{\"o}ttgens}, {Husser}, {Latour}, {Weilbacher}, \& {Wisotzki}}]{kamann2020a}
{Kamann}, S., {Giesers}, B., {Bastian}, N., {et~al.} 2020, \aap, 635, A65, \dodoi{10.1051/0004-6361/201936843}

\bibitem[{{Kim} \& {Lee}(2018)}]{kim2018}
{Kim}, J.~J., \& {Lee}, Y.-W. 2018, \apj, 869, 35, \dodoi{10.3847/1538-4357/aaec67}

\bibitem[{{Kravtsov} {et~al.}(2010){Kravtsov}, {Alca{\'\i}no}, {Marconi}, \& {Alvarado}}]{Kravtsov2010}
{Kravtsov}, V., {Alca{\'\i}no}, G., {Marconi}, G., \& {Alvarado}, F. 2010, \aap, 512, L6, \dodoi{10.1051/0004-6361/200913749}

\bibitem[{{Kravtsov} \& {Calder{\'o}n}(2021)}]{Kravtsov2021}
{Kravtsov}, V., \& {Calder{\'o}n}, F.~A. 2021, \aj, 161, 7, \dodoi{10.3847/1538-3881/abc423}

\bibitem[{{Kruijssen} {et~al.}(2019){Kruijssen}, {Pfeffer}, {Reina-Campos}, {Crain}, \& {Bastian}}]{kruijssen2019}
{Kruijssen}, J.~M.~D., {Pfeffer}, J.~L., {Reina-Campos}, M., {Crain}, R.~A., \& {Bastian}, N. 2019, \mnras, 486, 3180, \dodoi{10.1093/mnras/sty1609}

\bibitem[{{Lagioia} {et~al.}(2019){Lagioia}, {Milone}, {Marino}, \& {Dotter}}]{lagioia2019a}
{Lagioia}, E.~P., {Milone}, A.~P., {Marino}, A.~F., \& {Dotter}, A. 2019, \apj, 871, 140, \dodoi{10.3847/1538-4357/aaf729}

\bibitem[{{Lagioia} {et~al.}(2025){Lagioia}, {Milone}, {Legnardi}, {Cordoni}, {Dondoglio}, {Renzini}, {Tailo}, {Ziliotto}, {Carlos}, {Jang}, {Marino}, {Mohandasan}, {Qi}, {Rangwal}, {Bortolan}, \& {Muratore}}]{lagioia2025}
{Lagioia}, E.~P., {Milone}, A.~P., {Legnardi}, M.~V., {et~al.} 2025, The Astrophysical Journal, 979, 30, \dodoi{10.3847/1538-4357/ad98ee}

\bibitem[{{Lee}(2017)}]{lee2017a}
{Lee}, J.-W. 2017, \apj, 844, 77, \dodoi{10.3847/1538-4357/aa7b8c}

\bibitem[{{Lee} \& {Jang}(2016)}]{Lee2016}
{Lee}, Y.-W., \& {Jang}, S. 2016, \apj, 833, 236, \dodoi{10.3847/1538-4357/833/2/236}

\bibitem[{{Lee} {et~al.}(1999){Lee}, {Joo}, {Sohn}, {Rey}, {Lee}, \& {Walker}}]{lee1999}
{Lee}, Y.~W., {Joo}, J.~M., {Sohn}, Y.~J., {et~al.} 1999, \nat, 402, 55, \dodoi{10.1038/46985}

\bibitem[{{Lee} {et~al.}(2015){Lee}, {Joo}, \& {Chung}}]{lee2015}
{Lee}, Y.-W., {Joo}, S.-J., \& {Chung}, C. 2015, \mnras, 453, 3906, \dodoi{10.1093/mnras/stv1980}

\bibitem[{{Lee} {et~al.}(2005){Lee}, {Joo}, {Han}, {Chung}, {Ree}, {Sohn}, {Kim}, {Yoon}, {Yi}, \& {Demarque}}]{lee2005}
{Lee}, Y.-W., {Joo}, S.-J., {Han}, S.-I., {et~al.} 2005, \apjl, 621, L57, \dodoi{10.1086/428944}

\bibitem[{{Legnardi} {et~al.}(2022){Legnardi}, {Milone}, {Armillotta}, {Marino}, {Cordoni}, {Renzini}, {Vesperini}, {D'Antona}, {McKenzie}, {Yong}, {Dondoglio}, {Lagioia}, {Carlos}, {Tailo}, {Jang}, \& {Mohandasan}}]{legnardi2022}
{Legnardi}, M.~V., {Milone}, A.~P., {Armillotta}, L., {et~al.} 2022, \mnras, 513, 735, \dodoi{10.1093/mnras/stac734}

\bibitem[{{Legnardi} {et~al.}(2023){Legnardi}, {Milone}, {Cordoni}, {Lagioia}, {Dondoglio}, {Marino}, {Jang}, {Mohandasan}, \& {Ziliotto}}]{legnardi2023}
{Legnardi}, M.~V., {Milone}, A.~P., {Cordoni}, G., {et~al.} 2023, \mnras, 522, 367, \dodoi{10.1093/mnras/stad1056}

\bibitem[{{Leitinger} {et~al.}(2023){Leitinger}, {Baumgardt}, {Cabrera-Ziri}, {Hilker}, \& {Pancino}}]{leitinger2023}
{Leitinger}, E., {Baumgardt}, H., {Cabrera-Ziri}, I., {Hilker}, M., \& {Pancino}, E. 2023, \mnras, 520, 1456, \dodoi{10.1093/mnras/stad093}

\bibitem[{{Libralato} {et~al.}(2023){Libralato}, {Vesperini}, {Bellini}, {Milone}, {van der Marel}, {Piotto}, {Anderson}, {Aparicio}, {Barbuy}, {Bedin}, {Brown}, {Cassisi}, {Nardiello}, {Sarajedini}, \& {Scalco}}]{Libralato2023}
{Libralato}, M., {Vesperini}, E., {Bellini}, A., {et~al.} 2023, \apj, 944, 58, \dodoi{10.3847/1538-4357/acaec6}

\bibitem[{{Lucatello} {et~al.}(2015){Lucatello}, {Sollima}, {Gratton}, {Vesperini}, {D'Orazi}, {Carretta}, \& {Bragaglia}}]{lucatello2015}
{Lucatello}, S., {Sollima}, A., {Gratton}, R., {et~al.} 2015, \aap, 584, A52, \dodoi{10.1051/0004-6361/201526957}

\bibitem[{{Marino} {et~al.}(2008){Marino}, {Villanova}, {Piotto}, {Milone}, {Momany}, {Bedin}, \& {Medling}}]{marino2008a}
{Marino}, A.~F., {Villanova}, S., {Piotto}, G., {et~al.} 2008, \aap, 490, 625, \dodoi{10.1051/0004-6361:200810389}

\bibitem[{{Marino} {et~al.}(2024){Marino}, {Milone}, {Legnardi}, {Renzini}, {Dondoglio}, {Cavecchi}, {Cordoni}, {Dotter}, {Lagioia}, {Ziliotto}, {Bernizzoni}, {Bortolan}, {Carlos}, {Jang}, {Mohandasan}, {Muratore}, \& {Tailo}}]{marino2023}
{Marino}, A.~F., {Milone}, A.~P., {Legnardi}, M.~V., {et~al.} 2024, arXiv e-prints, arXiv:2401.06681, \dodoi{10.48550/arXiv.2401.06681}

\bibitem[{{Massari} {et~al.}(2019){Massari}, {Koppelman}, \& {Helmi}}]{massari2019}
{Massari}, D., {Koppelman}, H.~H., \& {Helmi}, A. 2019, \aap, 630, L4, \dodoi{10.1051/0004-6361/201936135}

\bibitem[{{Mehta} {et~al.}(2025){Mehta}, {Milone}, {Casagrande}, {Marino}, {Legnardi}, {Cordoni}, {Dondoglio}, {Jang}, {Lionetto}, {Ziliotto}, {Barbieri}, {Bernizzoni}, {Bortolan}, {Bouras Moreno Sanchez}, {Lagioia}, {Mohandasan}, \& {Muratore}}]{mehta2025}
{Mehta}, V.~J., {Milone}, A.~P., {Casagrande}, L., {et~al.} 2025, \mnras, 536, 1077, \dodoi{10.1093/mnras/stae2542}

\bibitem[{{Milone} \& {Marino}(2022)}]{milone2022a}
{Milone}, A.~P., \& {Marino}, A.~F. 2022, Universe, 8, 359, \dodoi{10.3390/universe8070359}

\bibitem[{{Milone} {et~al.}(2010){Milone}, {Piotto}, {King}, {Bedin}, {Anderson}, {Marino}, {Momany}, {Malavolta}, \& {Villanova}}]{milone2010a}
{Milone}, A.~P., {Piotto}, G., {King}, I.~R., {et~al.} 2010, \apj, 709, 1183, \dodoi{10.1088/0004-637X/709/2/1183}

\bibitem[{{Milone} {et~al.}(2012{\natexlab{a}}){Milone}, {Piotto}, {Bedin}, {King}, {Anderson}, {Marino}, {Bellini}, {Gratton}, {Renzini}, {Stetson}, {Cassisi}, {Aparicio}, {Bragaglia}, {Carretta}, {D'Antona}, {Di Criscienzo}, {Lucatello}, {Monelli}, \& {Pietrinferni}}]{milone2012b}
{Milone}, A.~P., {Piotto}, G., {Bedin}, L.~R., {et~al.} 2012{\natexlab{a}}, \apj, 744, 58, \dodoi{10.1088/0004-637X/744/1/58}

\bibitem[{{Milone} {et~al.}(2012{\natexlab{b}}){Milone}, {Piotto}, {Bedin}, {Aparicio}, {Anderson}, {Sarajedini}, {Marino}, {Moretti}, {Davies}, {Chaboyer}, {Dotter}, {Hempel}, {Mar{\'\i}n-Franch}, {Majewski}, {Paust}, {Reid}, {Rosenberg}, \& {Siegel}}]{milone2012a}
---. 2012{\natexlab{b}}, \aap, 540, A16, \dodoi{10.1051/0004-6361/201016384}

\bibitem[{{Milone} {et~al.}(2017){Milone}, {Piotto}, {Renzini}, {Marino}, {Bedin}, {Vesperini}, {D'Antona}, {Nardiello}, {Anderson}, {King}, {Yong}, {Bellini}, {Aparicio}, {Barbuy}, {Brown}, {Cassisi}, {Ortolani}, {Salaris}, {Sarajedini}, \& {van der Marel}}]{milone2017a}
{Milone}, A.~P., {Piotto}, G., {Renzini}, A., {et~al.} 2017, \mnras, 464, 3636, \dodoi{10.1093/mnras/stw2531}

\bibitem[{{Milone} {et~al.}(2020){Milone}, {Marino}, {Da Costa}, {Lagioia}, {D'Antona}, {Goudfrooij}, {Jerjen}, {Massari}, {Renzini}, {Yong}, {Baumgardt}, {Cordoni}, {Dondoglio}, {Li}, {Tailo}, {Asa'd}, \& {Ventura}}]{milone2020a}
{Milone}, A.~P., {Marino}, A.~F., {Da Costa}, G.~S., {et~al.} 2020, \mnras, 491, 515, \dodoi{10.1093/mnras/stz2999}

\bibitem[{{Milone} {et~al.}(2023{\natexlab{a}}){Milone}, {Cordoni}, {Marino}, {D'Antona}, {Bellini}, {Di Criscienzo}, {Dondoglio}, {Lagioia}, {Langer}, {Legnardi}, {Libralato}, {Baumgardt}, {Bettinelli}, {Cavecchi}, {de Grijs}, {Deng}, {Hastings}, {Li}, {Mohandasan}, {Renzini}, {Vesperini}, {Wang}, {Ziliotto}, {Carlos}, {Costa}, {Dell'Agli}, {Di Stefano}, {Jang}, {Martorano}, {Simioni}, {Tailo}, \& {Ventura}}]{milone2023a}
{Milone}, A.~P., {Cordoni}, G., {Marino}, A.~F., {et~al.} 2023{\natexlab{a}}, \aap, 672, A161, \dodoi{10.1051/0004-6361/202244798}

\bibitem[{{Milone} {et~al.}(2023{\natexlab{b}}){Milone}, {Marino}, {Dotter}, {Ziliotto}, {Dondoglio}, {Cordoni}, {Jang}, {Lagioia}, {Legnardi}, {Mohandasan}, {Tailo}, {Yong}, {Baimukhametova}, \& {Carlos}}]{milone2023b}
{Milone}, A.~P., {Marino}, A.~F., {Dotter}, A., {et~al.} 2023{\natexlab{b}}, \mnras, 522, 2429, \dodoi{10.1093/mnras/stad1041}

\bibitem[{{Mohandasan} {et~al.}(2024){Mohandasan}, {Milone}, {Cordoni}, {Dondoglio}, {Lagioia}, {Legnardi}, {Ziliotto}, {Jang}, {Marino}, \& {Carlos}}]{mohandasan2024}
{Mohandasan}, A., {Milone}, A.~P., {Cordoni}, G., {et~al.} 2024, \aap, 681, A42, \dodoi{10.1051/0004-6361/202347424}

\bibitem[{{Monelli} {et~al.}(2013){Monelli}, {Milone}, {Stetson}, {Marino}, {Cassisi}, {del Pino Molina}, {Salaris}, {Aparicio}, {Asplund}, {Grundahl}, {Piotto}, {Weiss}, {Carrera}, {Cebri{\'a}n}, {Murabito}, {Pietrinferni}, \& {Sbordone}}]{monelli2013}
{Monelli}, M., {Milone}, A.~P., {Stetson}, P.~B., {et~al.} 2013, \mnras, 431, 2126, \dodoi{10.1093/mnras/stt273}

\bibitem[{{Niederhofer} {et~al.}(2017){Niederhofer}, {Bastian}, {Kozhurina-Platais}, {Larsen}, {Hollyhead}, {Lardo}, {Cabrera-Ziri}, {Kacharov}, {Platais}, {Salaris}, {Cordero}, {Dalessandro}, {Geisler}, {Hilker}, {Li}, {Mackey}, \& {Mucciarelli}}]{niederhofer2017a}
{Niederhofer}, F., {Bastian}, N., {Kozhurina-Platais}, V., {et~al.} 2017, \mnras, 465, 4159, \dodoi{10.1093/mnras/stw3084}

\bibitem[{{Piotto} {et~al.}(2015){Piotto}, {Milone}, {Bedin}, {Anderson}, {King}, {Marino}, {Nardiello}, {Aparicio}, {Barbuy}, {Bellini}, {Brown}, {Cassisi}, {Cool}, {Cunial}, {Dalessandro}, {D'Antona}, {Ferraro}, {Hidalgo}, {Lanzoni}, {Monelli}, {Ortolani}, {Renzini}, {Salaris}, {Sarajedini}, {van der Marel}, {Vesperini}, \& {Zoccali}}]{piotto2015a}
{Piotto}, G., {Milone}, A.~P., {Bedin}, L.~R., {et~al.} 2015, \aj, 149, 91, \dodoi{10.1088/0004-6256/149/3/91}

\bibitem[{{Renzini} {et~al.}(2022){Renzini}, {Marino}, \& {Milone}}]{renzini2022a}
{Renzini}, A., {Marino}, A.~F., \& {Milone}, A.~P. 2022, \mnras, 513, 2111, \dodoi{10.1093/mnras/stac973}

\bibitem[{{Renzini} {et~al.}(2015){Renzini}, {D'Antona}, {Cassisi}, {King}, {Milone}, {Ventura}, {Anderson}, {Bedin}, {Bellini}, {Brown}, {Piotto}, {van der Marel}, {Barbuy}, {Dalessandro}, {Hidalgo}, {Marino}, {Ortolani}, {Salaris}, \& {Sarajedini}}]{renzini2015a}
{Renzini}, A., {D'Antona}, F., {Cassisi}, S., {et~al.} 2015, \mnras, 454, 4197, \dodoi{10.1093/mnras/stv2268}

\bibitem[{{Spitzer}(1987)}]{Spitzer1987}
{Spitzer}, L. 1987, {Dynamical evolution of globular clusters}

\bibitem[{{Stetson} {et~al.}(2019){Stetson}, {Pancino}, {Zocchi}, {Sanna}, \& {Monelli}}]{stetson2019}
{Stetson}, P.~B., {Pancino}, E., {Zocchi}, A., {Sanna}, N., \& {Monelli}, M. 2019, \mnras, 485, 3042, \dodoi{10.1093/mnras/stz585}

\bibitem[{{Vesperini} {et~al.}(2013){Vesperini}, {McMillan}, {D'Antona}, \& {D'Ercole}}]{vesperini2013a}
{Vesperini}, E., {McMillan}, S. L.~W., {D'Antona}, F., \& {D'Ercole}, A. 2013, \mnras, 429, 1913, \dodoi{10.1093/mnras/sts434}

\bibitem[{{Wang} {et~al.}(2020){Wang}, {Kroupa}, {Takahashi}, \& {Jerabkova}}]{wang2020}
{Wang}, L., {Kroupa}, P., {Takahashi}, K., \& {Jerabkova}, T. 2020, \mnras, 491, 440, \dodoi{10.1093/mnras/stz3033}

\bibitem[{{Yong} \& {Grundahl}(2008)}]{yong2008a}
{Yong}, D., \& {Grundahl}, F. 2008, \apjl, 672, L29, \dodoi{10.1086/525850}

\bibitem[{{Zennaro} {et~al.}(2019){Zennaro}, {Milone}, {Marino}, {Cordoni}, {Lagioia}, \& {Tailo}}]{zennaro2019a}
{Zennaro}, M., {Milone}, A.~P., {Marino}, A.~F., {et~al.} 2019, \mnras, 487, 3239, \dodoi{10.1093/mnras/stz1477}

\end{thebibliography}
\bibliographystyle{aasjournal}



\end{document}